\documentclass[%
 reprint,
preprintnumbers,
nofootinbib,
 amsmath,amssymb,
 aps,
 onecolumn,
floatfix,
]{revtex4-2}

\usepackage{graphicx}
\usepackage{dcolumn}
\usepackage{bm}

\usepackage[utf8]{inputenc}
\usepackage{hyperref}
\usepackage{cleveref}
\usepackage{booktabs}
\usepackage{nicefrac} 
\usepackage{physics}
\usepackage{siunitx}
\usepackage{tensor}
\usepackage{multirow}
\usepackage{subcaption}
\usepackage{xcolor}
\usepackage{tikz}
\usepackage{tikz-3dplot}
\usepackage[compat=1.1.0]{tikz-feynman}

\crefformat{section}{Sec.~#2#1#3}
\Crefformat{section}{Section~#2#1#3}

\crefformat{appendix}{App.~#2#1#3}
\Crefformat{appendix}{Appendix~#2#1#3}

\crefformat{table}{Tab.~#2#1#3}
\Crefformat{table}{Table~#2#1#3}

\crefformat{equation}{Eq.~(#2#1#3)}
\Crefformat{equation}{Equation~(#2#1#3)}
\crefrangeformat{equation}{Eqs.~(#3#1#4) to~(#5#2#6)}
\crefmultiformat{equation}{Eqs.~(#2#1#3)}{ and~(#2#1#3)}{, (#2#1#3)}{ and~(#2#1#3)}

\crefformat{figure}{Fig.~#2#1#3}

\begin{document}

\preprint{MS-TP-22-09, MITP-22-078}

\title{\boldmath The dipole formalism for massive initial-state particles and its application to dark matter calculations}

\author{J.~Harz}
\email{julia.harz@uni-mainz.de}
\affiliation{PRISMA$^+$ Cluster of Excellence \& Mainz Institute for Theoretical Physics,\\
Johannes Gutenberg University, 55099 Mainz, Germany}
\affiliation{Physik Department T70, Technische Universität München,\\
James-Franck-Straße 1, 85748 Garching, Germany}

\author{M.~Klasen}%
 \email{michael.klasen@uni-muenster.de}
 \affiliation{
	Institut f\"ur Theoretische Physik, Westf\"alische Wilhelms-Universit\"at M\"unster, Wilhelm-Klemm-Stra{\ss}e 9, D-48149 M\"unster, Germany
  }
\affiliation{School of Physics, The University of New South Wales, Sydney NSW 2052, Australia}

\author{M.Y.~Sassi}
 \email{mohamed.younes.sassi@desy.de}
 \affiliation{II. Institut für Theoretische Physik
Universität Hamburg,\\Luruper Chaussee 149, 22761 Hamburg, Germany
	}

 \author{L.P.~Wiggering}
 \email{luca.wiggering@uni-muenster.de}
 \affiliation{
	Institut f\"ur Theoretische Physik, Westf\"alische Wilhelms-Universit\"at M\"unster, Wilhelm-Klemm-Stra{\ss}e 9, D-48149 M\"unster, Germany
  }

\date{\today}

\begin{abstract}
The dark matter abundance plays a crucial role in the determination of the valid parameter space of models both in the case of a discovery of dark matter and in the context of exclusion limits. Reliable theoretical predictions of the dark matter relic density require technically demanding precision calculations, which were so far limited in their automation due to challenges in the treatment of infrared divergences appearing in higher order calculations. In particular, massive initial states need to be considered in early Universe computations, so that the known dipole subtraction methods could not be directly exploited. We therefore provide a full generalization of the dipole subtraction method by Catani and Seymour to supersymmetric (SUSY) QCD with massive initial states. All dipole splitting functions and their integrated counterparts are given explicitly for four different dimensional schemes. To showcase their application, we apply our results to dark matter (co)annihilation processes in the context of the minimal supersymmetric Standard Model. We also demonstrate the accuracy of the dipole method by comparing our numerical results with those obtained with the phase space slicing method. Our analytical results will facilitate future automation of dark matter abundance calculations at next-to-leading order for both SUSY and non-SUSY models.
\end{abstract}

\maketitle


\section{Introduction}
The very precise measurement of the present amount of dark matter in the Universe by the Planck satellite allows to place stringent constraints on dark matter models \cite{Planck:2018vyg}. In order to keep up with the experimental uncertainty, next-to-leading order (NLO) corrections have to be included in theoretical calculations of the relic abundance \cite{Baro:2007em,Herrmann:2009mp,Harz:2012fz,Harz:2014tma,Harz:2014gaa,Harz:2016dql,Schmiemann:2019czm,Branahl:2019yot}. The associated numerical evaluation of real emission processes is problematic in phase space regions where the squared matrix element becomes soft or collinear, as only the sum of the real and virtual corrections is infrared finite. The two main general approaches which allow the analytic cancellation of infrared singularities between both contributions are subtraction methods 
\cite{Frixione:1995ms,Catani:1996vz,Catani:2002} and the phase space slicing (PSS) method \cite{Giele:1991vf,Fabricius:1981sx,Kramer:1986mc,Harris:2001sx}. A general treatment of massive initial particles, e.g.\ in supersymmetric (SUSY) QCD, as required for dark matter (co)-annihilation processes is available for the slicing approach, but not for the subtraction methods. Dittmaier considered photon radiation off heavy fermions in QED, but used a small photon mass as a regulator \cite{Dittmaier:1999mb}. Consequently, the results cannot be simply transferred to QCD, where divergences are commonly regularized via a dimensional scheme. Kotko \cite{Kotko:2012agp, Kotko:2012ui} previously considered a fully massive dipole formalism for initial-final dipoles in conventional dimensional regularization, leaving out e.g. the emission off a massless final-state quark with a massive spectator and in a different convention of parameters compared to the one used in this work, complicating the computation of necessary integration limits. Ref. \cite{Krauss:2017wmx} focuses primarily on the case of the initial-initial dipole configuration corresponding to the emission of a gluon into the final state off a massive initial quark and pays particular attention to the necessary modifications of the standard treatment of parton distribution functions in the final dipole formulae required by the inclusion of mass effects. The phase space slicing method has been successfully applied to dark matter calculations in the past \cite{Harz:2012fz,Harz:2014tma,Schmiemann:2019czm}, but this approach has the practical disadvantages that the squared real emission matrix element has to be subdivided into finite, soft, collinear and soft-collinear contributions and that the final result depends on the chosen cutoffs. In addition, the slicing method is found to be less accurate and efficient compared to the dipole approach \cite{Eynck:2001en}.

For these reasons, it is the objective of this paper to extend the Catani-Seymour dipole subtraction method to massive initial states for initial-final as well as initial-initial dipoles in a \emph{unified} notation similar to the one used in Ref. \cite{Catani:2002} and provide all formulae for squark and gluino (co)-annihilation as required by dark matter calculations. We also pay particular attention to provide all formulae for different dimensional schemes, as conventional dimensional regularization breaks supersymmetry already at the one-loop level in contrast to dimensional reduction, which is therefore the preferred scheme for calculations in supersymmetry.
By a simple change of the color factor the results can also be applied to heavy scalar and fermionic dark matter in general. The provided formulae also allow for one massive and one massless particle in the initial state. The results do not apply to processes with identified (R)-hadrons and to splitting processes where the mass of the parent particle is unequal to the mass of one of its decay products such as the splitting $g\to q\bar{q}$ into massive quarks.

The paper is organized in the following way: in \cref{sec:review} we review the dipole subtraction method for the case of no (R)-hadrons in the initial or final state. We also cover the factorization of (SUSY)-QCD amplitudes in the soft and (quasi)-collinear limit for the construction of the dipole splitting functions. The main part from \cref{sec:FEIS} to \cref{sec:FEFS} provides the dipole splitting functions along with the integrated counterparts and a detailed account of the integration technique for the three possible emitter-spectator pairs with at least one colored initial state. \Cref{sec:comparision} covers the application of the dipole method to the example processes $\tilde{\chi}_1^0 \tilde{t}_1\to t g$ and $\tilde{t}_1 \tilde{t}_1\to t t$ and the corresponding comparison with the phase space slicing method. Our summary is given in \cref{sec:summary}. In the appendix, further details on the phase space factorization are provided and the computation of the required non-trivial integrals is sketched. In addition, we define precisely the four different dimensional schemes that we distinguish in our calculation for a better common understanding. 

\section{Review of the dipole subtraction method}
\label{sec:review}
A generic cross section $\sigma^{\text{NLO}}$ describing the production of $m$ particles at next-to-leading order (NLO) accuracy in (SUSY)-QCD without initial-state (R)-hadrons can be decomposed as
\begin{equation}
\sigma^{\text{NLO}} = \sigma^{\text{Tree}}+\Delta \sigma^{\text{NLO}} =  \int \dd\sigma^{\text{B}}+\Delta \sigma^{\text{NLO}},
\end{equation} 
where $\dd\sigma^{\text{B}}$ denotes the differential tree-level cross section and the NLO part $\Delta \sigma^{\text{NLO}}$ receives contributions from virtual corrections $\dd\sigma^{\text{V}}$ as well as from real emission $\dd\sigma^{\text{R}}$ of massless particles 
\begin{equation}
\Delta \sigma^{\text{NLO}} = \int_m \dd\sigma^{\text{V}} + \int_{m+1} \dd\sigma^{\text{R}}.
\label{eq:NLO_Xsec}
\end{equation}
The subscript on the integrals refers to the number of particles in the final state. The inclusion of (R)-hadrons would require a proper factorization of short and long distance physics whereas we assume that the cross section is perturbative. 
After successful renormalization the virtual part is ultraviolet finite, but still contains another type of divergence: the infrared (IR) divergence which appears when the loop momentum of a massless virtual particle becomes almost zero or collinear to the direction of another massless particle. Therefore, one distinguishes between soft, collinear and soft-collinear infrared divergences, where in the latter case the massless particle is soft and collinear at the same time.

The same kind of infrared behavior occurs within the real contribution. According to the Kinoshita-Lee-Nauenberg theorem \cite{Kinoshita:1962ur}, every unitary quantum field theory such as the Standard Model or its minimal supersymmetric extension [the minimal supersymmetric Standard Model (MSSM)] is infrared finite as a whole. As a consequence, the IR divergences from the phase space integration of the real part cancel those coming from loop integrals of the virtual part on the right hand side of \cref{eq:NLO_Xsec}. In practice, the divergences have to be extracted with the help of a regulator such as an artificial mass. However, the only known regularization procedure which preserves gauge and Lorentz invariance (as well as supersymmetry) is dimensional regularization (dimensional reduction). Within these procedures, the number of space-time dimensions is continued analytically from four to $D=4-2\varepsilon$. In this regularization scheme, soft and collinear divergences take the form of simple poles in $\varepsilon$, whereas soft-collinear divergences appear as double poles. 
Due to the large number of terms that enter during the standard Feynman-diagrammatic calculation of (SUSY)-QCD matrix elements, it is often impossible to perform the integration over the $m+1$ particle phase space in \cref{eq:NLO_Xsec} analytically in $D$ dimensions except for the very simplest processes. In order to make a numerical evaluation of the real emission matrix elements over the whole phase space possible, \textbf{}i.e. without relying on cuts and approximations as in the phase space slicing approach, Catani and Seymour developed the dipole subtraction method \cite{Catani:1996vz}. The basic idea is to construct an auxiliary cross section $\dd\sigma^{\text{A}}$ which converges pointwise to $\dd\sigma^{\text{R}}$ in the singular region in $D$ dimensions, so that $\dd{\sigma}^{\text{R}}-\dd\sigma^{\text{A}}$ is finite over the whole region of phase space and can be integrated in four dimensions. At the same time it must be possible to integrate $\dd\sigma^{\text{A}}$ analytically in $D$ dimensions over the one-particle phase space of the radiated massless particle giving rise to the divergence. This allows to add back the subtraction term and to cancel those divergences appearing in the virtual contribution which are present in the form of simple or double poles in $\varepsilon$. The computation of the NLO correction can then be summarized as
\begin{equation}
\Delta\sigma^{\text{NLO}}=\int_{m+1}\left[\dd{\sigma}^{\text{R}}_{\varepsilon=0}-\dd\sigma^{\text{A}}_{\varepsilon=0}\right]+\int_{m}\left[\dd{\sigma}^{\text{V}}+\int_{1}\dd\sigma^{\text{A}}\right]_{\varepsilon=0}.
\end{equation}
The counterterm $\dd\sigma^{\text{A}}$ is constructed from the knowledge that QCD amplitudes factorize in the soft and collinear limit in the process-dependent Born level cross section $\dd\sigma^{\text{B}}$ convolved with a universal splitting kernel $\dd\mathbf{V}_{\text{dipole}}$, which reflects the singular behavior. From another point of view, the factorization can be thought of as a two-step process. In the first step, $m$ final state particles are produced through the Born level cross section $\dd\sigma^{\text{B}}$. In the second step, the final $(m+1)$-particle configuration is reached through the decay of one of the $m$ particles - \emph{the emitter} - into two particles. This last step is described by the splitting function $\dd \mathbf{V}_{\text{dipole}}$. The information about color and spin correlations is accounted for by referencing an additional particle - \emph{the spectator}. The final expression for $\dd\sigma^{\text{A}}$ is obtained by summing over all possible emitter-spectator pairs 
\begin{equation}
     \int_{m+1}\dd\sigma^{\text{A}}=\sum_{\text{dipoles}} \int_{m}\dd\sigma^{\text{B}} \otimes \int_{1}\dd\mathbf{V}_{\text{dipole}} = \sum_{\text{dipoles}} \int_{m}\left[\dd\sigma^{\text{B}} \otimes \mathbf{I}\right] , \label{eq:defI}
\end{equation}
where the universal factor $\mathbf{I}$ corresponds to the integral of the dipole splitting function over the one particle phase space, and thus cancels the infrared divergences in the virtual part.
The fact that the underlying structure of this factorization is formed by these pairs lead to the name "dipole formalism". However, as this factorization holds only in the strict soft and collinear limit and it is desirable that $\dd{\sigma^{\text{A}}}$ approximates $\dd{\sigma}^{\text{R}}$ also in a small region around the singularity to render the subtraction procedure numerically stable, one has to introduce the so-called \emph{dipole momenta} to ensure that the factorization does not violate momentum conservation. These obey momentum conservation in the whole $m+1$-particle phase space and are defined through a smooth map from the $m+3$ real emission momenta to the $m+2$ dipole momenta. Their precise definition depends on the kinematical situation and therefore their concrete expressions will be given in the sections dedicated to the different emitter-spectator pairs.

In order to allow for a general construction of the auxiliary cross section, the aforementioned color and spin correlations are implemented into the factorization formula by realizing the splitting functions $\mathbf{V}_{\text{dipole}}$ as operators that act on matrix elements which are defined as abstract objects in color and spin space. For this purpose we make use of the of conventions and the notation established in Refs. \cite{Catani:1996vz,Catani:2002} which we introduce in the following. That is, colored particles in the initial state are labelled by $a,b,\dots$ and those in the final state by $i,j,k,\dots$. Since non-colored particles are irrelevant for the subtraction procedure, they are suppressed in the notation. Scattering amplitudes are considered as objects in an abstract vector space spanned by the spins $s_a, s_i$ and colors $c_a, c_i$ of all colored particles involved in the process
\begin{equation}
\ket{\{i,a\}}_m  = \frac{1}{\prod_b \sqrt{n_c(b)}}\mathcal{M}_m^{\{c_i, s_i; c_a,s_a\}}\left(\left\{p_i;p_a\right\}\right) \left(\ket{\left\{c_i;c_a\right\}}\otimes \ket{\left\{s_i; s_a\right\}}\right) \label{eq:M2_as_ket}
\end{equation}
where $\prod_b \sqrt{n_c(b)}$ fixes the normalization by averaging over the $n_c(b)$ color degrees of freedom for each initial particle $b$.
The kets $\ket{\left\{c_i;c_a\right\}}$ and $\ket{\left\{s_i; s_a\right\}}$ constitute formally an orthogonal basis of the color and spin space, respectively. The color charge operators $\mathbf{T}_i$ or $\mathbf{T}_a$ reflect the emission of a gluon (or another massless colored particle) from a particle $i$ or $a$. Their action on color space is defined as 
\begin{multline} 
\tensor*[_m]{\bra{\{i,a\}}}{}\mathbf{T}_j\cdot \mathbf{T}_{k} \ket{\{i,a\}}_m = \frac{1}{\prod_b n_c(b)}\left[\mathcal{M}_m^{c_1,\dots,c_j,\dots,c_k,\dots,c_m;\{a\}}\left(\left\{p_i;p_a\right\}\right)\right]^{\ast}  \\ 
\times \mathcal{T}^e_{c_j d_j} \mathcal{T}^e_{c_k d_k} \mathcal{M}_m^{d_1,\dots,d_j,\dots,d_k,\dots,d_m;\{a\}}\left(\left\{p_i;p_a\right\}\right) \label{eq:action_color_operator}
\end{multline}
and analogously if $j$ or $k$ are initial-state particles. For a final-state particle $j$, the color charge matrix $\mathcal{T}^e_{c d}$ is defined as 
\begin{align}
    \mathcal{T}^e_{c_j d_j}=\begin{cases}
    -i f_{c_j d_j e}\\ 
     T^e_{c_j d_j}  \\
     -T^e_{d_j c_j} 
    \end{cases}
    \text{if $j$ is in the}  \ \ \ \begin{aligned}
     &\text{adjoint} \\ 
     &\text{fundamental} \\
     &\text{anti-fundamental}
    \end{aligned} \ \ \
    \text{ representation of $\mathfrak{su}(3)_c$},
\end{align}
with $T^a=\frac{\lambda^a}{2}$ being half of the Gell-Mann matrices $\lambda^a$ and $f_{abc}$ the structure constants of $\mathfrak{su}(3)_c$. The color charge operator $\mathbf{T}_a$ of an initial particle $a$ obeys the same action defined in \cref{eq:action_color_operator}. However, by crossing symmetry the color charge matrix in this case is defined as 
\begin{align}
    \mathcal{T}^e_{c_a d_a}=\begin{cases}
    -i f_{c_a d_a e}\\ 
     - T^e_{d_a c_a}  \\
    T^e_{c_a d_a}
    \end{cases}
    \text{if $a$ is in the}  \ \ \ \begin{aligned}
     &\text{adjoint} \\ 
     &\text{fundamental} \\
     &\text{anti-fundamental}
    \end{aligned} \ \ \
    \text{ representation of $\mathfrak{su}(3)_c$}.
\end{align}
Since each ket $\ket{\{i,a\}}_m$ must be a color singlet, color conservation can be written as 
\begin{equation}
\left(\sum_{j} \mathbf{T}_j + \sum_{b} \mathbf{T}_b \right) \ket{\{i,a\}}_m = \sum_{I} \mathbf{T}_I \ket{\{i,a\}}_m = 0,
\end{equation}
where we introduced the index $I$ which runs over both initial and final state particles. Furthermore, the commutation relation
\begin{equation}
    \comm{\mathbf{T}_i}{\mathbf{T}_j}=0 \ \text{if $i\neq j$}, \ \ \mathbf{T}_i^2 = C_i = \begin{cases} C_A, \ \text{$i$ adjoint} \\ C_F, \ \text{$i$ (anti)-fundamental} \end{cases} 
\end{equation}
with the quadratic Casimir operators $C_i$ follows directly from the definition of the color charge operators.

With these definitions and conventions at hand, we can move on to the explicit construction of the dipole splitting functions which approximate the real emission matrix element in the soft and collinear limit. 
\emph{In the soft limit}, where the momentum of a gluon $i$ tends to zero, the real emission matrix element can be written in terms of an eikonal current of the gluon
\begin{equation}
    \mathbf{J}^\mu = \sum_a \frac{p_a^\mu}{p_a \cdot p_i}  \mathbf{T}_a+ \sum_j \frac{p_j^\mu}{p_j \cdot p_i} \mathbf{T}_j = \sum_{I} \frac{p_I^\mu}{p_I \cdot p_i}  \mathbf{T}_I
\end{equation} 
and behaves as
\begin{multline}
   \tensor*[_{m+1,a\dots}]{\braket{\dots,i,\dots,j,\dots;a,\dots}{\dots,i,\dots,j,\dots;a,\dots}}{_{m+1,a\dots}}  \ \overset{p_i\to 0}{\xrightarrow{\hspace*{0.9cm}}} \\ -4\pi \mu^{2\varepsilon}\alpha_s \tensor*[_{m,a\dots}]{\bra{\dots,j,\dots;a,\dots}}{} \mathbf{J}^\dag_\mu \mathbf{J}^\mu\ket{\dots,j,\dots;a,\dots}_{m,a\dots} 
\end{multline}
with the strong coupling $\alpha_s$. The renormalization scale $\mu$ comes from the transition from four to $D$ space-time dimensions and ensures that the strong coupling remains nondimensional.
By using partial fractioning 
\begin{align}
 \frac{p_I \cdot p_K}{(p_I\cdot p_i)(p_K\cdot p_i)} =  \frac{p_I \cdot p_K}{(p_I\cdot p_i)(p_I+p_K)\cdot p_i} + \frac{p_I \cdot p_K}{(p_K\cdot p_i)(p_I+p_K)\cdot p_i}
\end{align}
and color conservation, the squared eikonal current can be recast into a sum over emitter ($I$) and spectator ($K$) pairs
\begin{equation}
    \mathbf{J}^\dag_\mu \mathbf{J}^\mu = \sum_{I,K} \frac{p_I\cdot p_K}{(p_I \cdot p_i)(p_K\cdot p_i)} \mathbf{T}_I \cdot \mathbf{T}_K \sum_{
    \substack{I,K \\ I\neq K }} \frac{1}{p_I\cdot p_i} \left(\frac{2 p_I \cdot p_K}{(p_I+p_K)\cdot p_i}-\frac{m_I^2}{p_I\cdot p_i}\right) \mathbf{T}_I \cdot \mathbf{T}_K.
\end{equation}
\begin{figure}[h]
    \centering
    \includegraphics{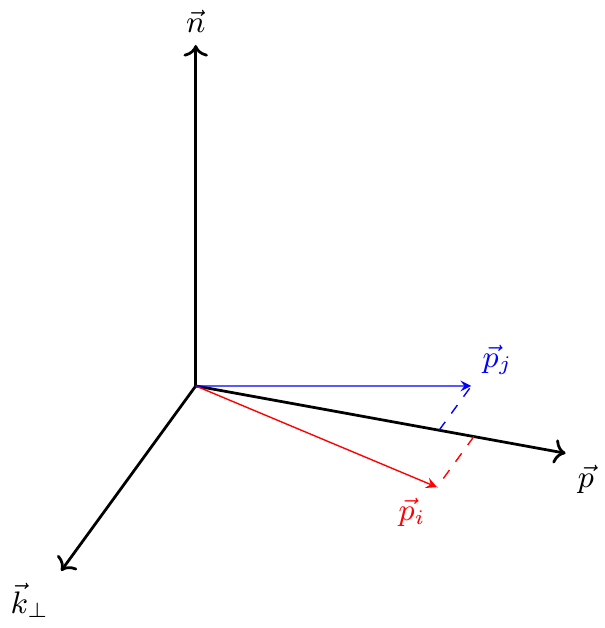}
     
     
     
      
      
     
    \caption{Sudakov vector parametrization.}
    \label{fig:sudakov}
\end{figure}
For two final-state particles $i$ and $j$ that are produced through a splitting $\widetilde{ij}\to i +j$ of a parent particle $\widetilde{ij}$, there is also a \emph{collinear divergence} if $i$ and $j$ are massless or a \emph{quasi-collinear} divergence if $i$ and $j$ are massive but their mass is small compared to the energy scale of the calculation so that the true collinear divergence is screened by the non-zero mass. In order to make the divergence visible, their momenta $p_i$ and $p_j$ can be expressed through the Sudakov parametrization 
\begin{align}
    &p_i^\mu = z p^\mu + k_\perp^\mu - \frac{k_\perp^2+z^2 m^2_{ij}-m_i^2}{z} \frac{n^\mu}{2 p\cdot n} , \\
    &p_j^\mu = (1-z) p^\mu -k_\perp^\mu -\frac{k_\perp^2+(1-z)^2 m^2_{ij}-m_j^2}{1-z} \frac{n^\mu}{2 p\cdot n} ,
\end{align}
where the timelike momentum $p$ with $p=m^2_{ij}$ gives the collinear direction and an auxiliary lightlike four-vector $n$ is needed to specify the transverse component $k_\perp$ which is perpendicular to $n$ and $p$ ($k_\perp\cdot n  =k_\perp\cdot p =0$), cf.\ Fig.\ \ref{fig:sudakov}. The variable $z$ corresponds to the momentum fraction involved in the splitting. With the help of this parametrization, the squared real emission matrix element reduces in the (quasi)-collinear limit to
\begin{multline}
   \tensor*[_{m+1,a\dots}]{\braket{\dots,i,j,\dots;a,\dots}{\dots,i,j,\dots;a,\dots}}{_{m+1,a\dots}} \ \overset{p_i\parallel p_j}{\xrightarrow{\hspace*{0.9cm}} } 
   \\ \frac{4\pi \mu^{2\varepsilon}\alpha_s}{p_i\cdot p_j}  \tensor*[_{m,a\dots}]{\bra{\dots,\widetilde{ij},\dots;a,\dots}}{}\hat{P}_{\widetilde{ij},i}(z,k_\perp;\varepsilon)\ket{\dots,\widetilde{ij},\dots;a,\dots}_{m,a\dots} 
\end{multline}
with the (generalized) Altarelli-Parisi \cite{ALTARELLI1977298} splitting function $\hat{P}_{\widetilde{ij},i}(z,k_\perp;\varepsilon)$. For the process $q \to q+g$ we are interested in the cases of massless as well as massive quarks and include for this reason the quasi-collinear limit which corresponds to the collinear one in the zero-mass limit.  We only consider the pure collinear limit for the splittings $g \to q + \bar{q}$ and $g\to g g$. The associated splitting functions are given by \cite{ALTARELLI1977298,Catani:1996pk,Catani:2002}
\begin{subequations}
\label{eq:altarelli-parisi}
\begin{align}
    \bra{s}\hat{P}_{qg}(z,k_\perp;\varepsilon)\ket{s'} &=  \delta_{s s'} C_F \left[\frac{2 (1-z)}{z} + \frac{1}{2} h_g^{\textsc{RS}} z-\frac{m^2_q}{p_g\cdot p_q}\right], \\
    \bra{\mu}\hat{P}_{gq}(z,k_\perp;\varepsilon)\ket{\nu} &= T_F \left[-g^{\mu\nu} + 4 z (1-z) \frac{k^\mu_\perp k^\nu_\perp}{k^2_\perp}\right], \\
    \bra{\mu}\hat{P}_{gg}(z,k_\perp;\varepsilon)\ket{\nu} &= 2 C_A \left[-g^{\mu\nu} \left(\frac{z}{1-z}+\frac{1-z}{z}\right)- h_g^{\textsc{RS}} z (1-z) \frac{k^\mu_\perp k^\nu_\perp}{k^2_\perp}\right].
\end{align}
\end{subequations}
The number of internal helicity states of the gluon $h_g^{\textsc{RS}}$ are introduced in \cref{eq:altarelli-parisi} to distinguish between different variants of dimensional regularization. Its precise definition for the four different dimensional schemes that we distinguish as well as the definition of the schemes themselves are provided in \cref{sec:dimension}.
For the construction of the dipole splitting function $\mathbf{V}_{\text{dipole}}$, we need to take into account both, the soft and collinear limit. However, it is not simply possible to add both limits as this will lead to an "over-counting" of the soft divergence, as the Altarelli-Parisi splitting functions also diverge in the soft limit. Therefore, it is necessary to construct the dipole splitting functions such that both limits are fulfilled separately, i.e. the overlapping region is only taken into account once.

The final \emph{dipole factorization formula} that defines the auxiliary squared matrix element related to $\dd{\sigma^{\text{A}}}$ is
\begin{equation}
\left|\mathcal{M}^{\text{A}}_{ m+1}\right|^2 = \sum_{i,j}\sum_{k\neq i,j} \mathcal{D}_{ij,k}+\sum_{i,j}\sum_{a} \mathcal{D}^a_{ij}+\sum_{a,i}\sum_{j\neq i} \mathcal{D}^{ai}_{j}+\sum_{a,i}\sum_{b\neq a} \mathcal{D}^{ai,b},
\label{eq:dipole_formula}
\end{equation}
where one has to distinguish between four different dipoles for the four different initial/final-state combinations of emitter and spectator. The precise definition of the dipoles $\mathcal{D}^a_{ij}$, $\mathcal{D}^{ai}_{j}$ and $\mathcal{D}^{ai,b}$ related to the splitting kernels $\mathbf{V}_{\text{dipole}}$ as well as the process dependent kernels themselves will be given in the following sections. We will not provide a definition for the dipole $\mathcal{D}_{ij,k}$ where emitter and spectator are both from the final state as this case is already fully covered for the massive and the massless case in Refs. \cite{Catani:1996vz,Catani:2002}.

\section{Final-state emitter and initial-state spectator}
\label{sec:FEIS}
The dipole contribution $\mathcal{D}_{ij}^a$ in \cref{eq:dipole_formula} is defined as
\begin{equation}
\mathcal{D}_{ij}^a=\frac{1}{-2 p_i\cdot p_j} \frac{1}{x_{ij,a}}\tensor*[_{m,a}]{\bra{\dots,\widetilde{ij},\dots;\tilde{a},\dots}}{}\frac{\mathbf{T}_a\cdot \mathbf{T}_{ij}}{\mathbf{T}_{ij}^2} \mathbf{V}_{ij}^a\ket{\dots,\widetilde{ij},\dots;\tilde{a},\dots}_{m,a},
\label{eq:dipole_FEIS}
\end{equation}
where the function $\mathbf{V}_{ij}^a$ describes the splitting process $\widetilde{ij} \to i+j$. The variable $x_{ij,a}$ will be defined in the section on the kinematical quantities used for the formulation of the splitting kernels. The tree matrix element with $m$ final-state particles is obtained from the original one with $(m+1)$ particles by replacing $i$ and $j$ with the emitter $\widetilde{ij}$ of momentum $\tilde{p}_{ij}$ and by exchanging the initial particle $a$ with $\tilde{a}$ of momentum $\tilde{p}_a$. In the following, we consider only the specific case $m_{ij}=m_j$ where the mass of the emitter $\widetilde{ij}$ is identical to the one of $j$ as the more general case $m_{ij}\neq m_j$ case is not needed for the example processes. 

Since a treatment of massless initial particles is already available in the literature \cite{Catani:1996vz,Catani:2002}, the initial particle $a$ will be treated as massive throughout this paper, whereas the final-state particle with momentum $p_j$ has an arbitrary mass and the mass of $i$ is zero,
\begin{equation}
p_a^2=m_a^2>0, \ \ \ \ \ \ \ \ \ \ p_j^2=m_j^2, \ \ \ \ \ \ \ \ \ \ p_i^2=0.
\end{equation}

\subsection{Kinematics} \label{sec:FEIS_kinematics}

\begin{figure}
    \centering
    \includegraphics{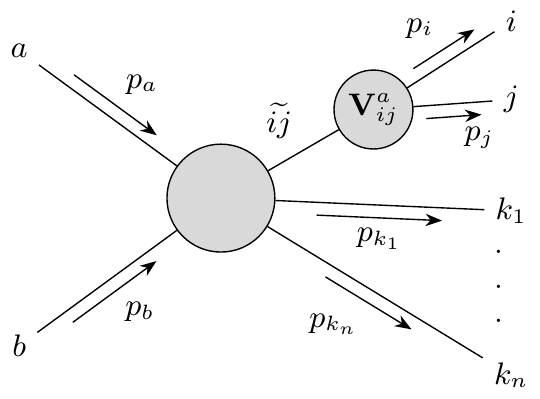}

    \caption{Diagrammatic interpretation of the dipole $\mathcal{D}^a_{ij}$ and the associated splitting function $\mathbf{V}^a_{ij}.$}
    \label{fig:3legs}
\end{figure}

\begin{figure}
    \centering
\includegraphics{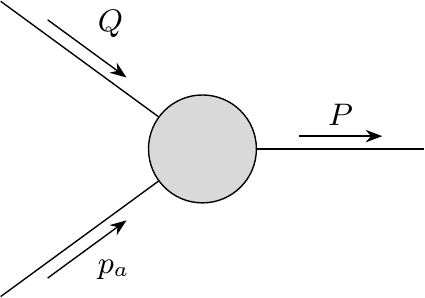}
\hspace{2cm}
\includegraphics{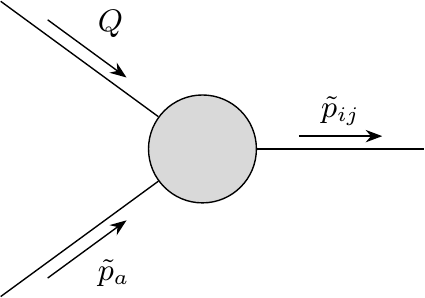}
    \caption{Kinematics for a final-state emitter and an initial-state spectator in the original momenta (left) and the dipole momenta (right).}
    \label{fig:FEIS-dipoleKinematics}
\end{figure}

For the construction of the dipole $\mathcal{D}_{ij}^a$ we adopt the kinematic quantities introduced in Ref. \cite{Dittmaier:1999mb} for photon emission off massive fermions. However, in that paper a small photon mass is used as infrared regulator as it is common in electroweak physics, so that the crucial part lies in the generalization of the phase space parametrization from four to $D$ dimensions.  

The two main quantities are the total outgoing momentum of the dipole phase space
\begin{equation} 
P=p_i+p_j
\end{equation}
and the total transferred momentum
\begin{align}
Q&=P-p_a=p_b-\sum_k p_k=\tilde{p}_{ij} - \tilde{p}_a,
\end{align}
where $k$ runs over the momenta of all other $(m-1)$ final-state particles besides $p_i$ and $p_j$, cf.\ Figs.\ \ref{fig:3legs} and \ref{fig:FEIS-dipoleKinematics}. At this point, one should highlight the difference between $P$ and $\widetilde{p}_{ij}$. That is, $P$ is the true momentum of the parent particle $\tilde{ij}$ in the real emission matrix element
whereas $\tilde{p}_{ij}$ is the dipole momentum which is inserted into the tree matrix element as momentum of $\widetilde{ij}$ within the auxiliary matrix element. Before we can define the dipole splitting functions and the dipole momenta explicitly, some auxiliary variables have to be introduced. These are first of all the momentum fractions
\begin{align}
z_j=\frac{p_a\cdot p_j}{P\cdot p_a} =1-z_i,   \ \ \ \ \ \ \ 
x_{ij,a}=\frac{P\cdot p_a-p_i\cdot p_j}{P\cdot p_a},
\end{align}
which take by definition only values between zero and one and behave in the soft $\left(p_i^\mu \to 0\right)$ and collinear limit $\left(p_i\cdot p_j\to 0\right)$  as
\begin{equation}
z_i \to 0, \ \ \ \  z_j \to 1, \ \ \ \ x_{ij,a}\to 1.
\end{equation}
The different quantities are related through
\begin{align}
P^2=\frac{-\bar{Q}^2}{x_{ij,a}}+Q^2-m_a^2 \label{eq:relation_P2Q2x}, \ \ \ \ \ \ \ \ \   \ \ \ 
P\cdot p_a=\frac{-\bar{Q}^2}{2x_{ij,a}} ,
\end{align}
where we introduced the abbreviation 
\begin{equation}
    \bar{Q}^2=Q^2-m_a^2-m_j^2.
\end{equation}
It is worth noting that since the product $P\cdot p_a$ is always positive and $x_{ij,a}$ can only take values between zero and one, $\bar{Q}^2$ is always negative such that $\sqrt{\bar{Q}^4}=-\bar{Q}^2$ with $\bar{Q}^4=\left(\bar{Q}^2\right)^2$. In addition, we define the auxiliary variables 
\begin{align}
\lambda_{aj}&=\lambda\left(Q^2,m_j^2,m_a^2\right)=\bar{Q}^4-4 m_a^2 m_j^2 \ \ \ \ \text{and} \\
R(x)&=\frac{\sqrt{\left(\bar{Q}^2+2 m_a^2 x\right)^2 - 4 m_a^2 Q^2 x^2}}{\sqrt{\lambda_{aj}}} \label{eq:Raj_var}
\end{align}
with the Källén function 
\begin{equation}
\lambda(x,y,z) =  x^{2}+y^{2}+z^{2}-2xy-2yz-2zx
\end{equation}
as well as the reduced masses $\eta_n$ and the relativistic relative velocity $v$ between $\tilde{p}_{ij}$ and $\tilde{p}_{a}$
\begin{equation}
    \eta_n = \frac{m^2_n}{-\bar{Q}^2} \ \ (n=a,j), \ \ \ \ \ \ \ \ \ \
     v=\frac{\sqrt{\lambda_{aj}}}{-\bar{Q}^2} \label{eq:v_and_eta}.
\end{equation}
It is straightforward to check that $P^2\to m_j^2$ and $R(x_{ij,a})\to 1$ in the soft and collinear limit.
The dipole momenta of emitter and spectator 
\begin{align}
\tilde{p}_{ij}^\mu&=\frac{x_{ij,a}}{R(x_{ij,a})} p_a^{\mu} + \left(\frac{1}{R(x_{ij,a})} \frac{\bar{Q}^2 + 2 m_a^2 x_{ij,a}}{2 Q^2} - \frac{Q^2 + m_a^2 - m_j^2}{2 Q^2} + 1\right) Q^\mu \label{eq:dip_momentum_pij}\\
\tilde{p}_a^\mu &= \tilde{p}_{ij}^\mu - Q^\mu \label{eq:dip_momentum_pa}
\end{align}
are constructed from the requirement to fulfill the on-shell conditions $\tilde{p}_a^2= m_a^2$, $\tilde{p}_{ij}^2=m_{ij}^2$ and momentum conservation $\tilde{p}_a+p_b= \tilde{p}_{ij} + p_k$.

\subsection{Phase space factorization}
The factorization of the $(m+1)$-particle phase space  $\dd\phi_{m+1}\left(p_i,p_j,p_k;p_a+p_b\right)$ into the $m$-particle phase space $\dd\phi_m\left(P(x),p_k;p_a+p_b\right)$ and the dipole phase space $\left[\dd p_i\left(Q^2,x,z_i\right)\right]$ is derived in \cref{sec:PSFE-IS} and corresponds to a convolution over the parameter $x$ 
\begin{equation}
 \int \dd\phi_{m+1}\left(p_i,p_j,p_k;p_a+p_b\right) \theta(x_{ij,a}-x_0)  =\int_{x_0}^{1} \dd{x} \int \dd\phi_m\left(P(x),p_k;p_a+p_b\right) \int \left[\dd p_i\left(Q^2,x,z_i\right)\right], \label{eq:PSfactorizFEIS}  
\end{equation}
where $x$ plays the role of $x_{ij,a}$. In \cref{eq:PSfactorizFEIS} an additional auxiliary parameter $x_0$ with $0\leq x_0 < 1$ is introduced as a lower limit on $x$ which is provided by the constraint that the argument of the square root in \cref{eq:Raj_var} remains positive for all possible values of $Q^2$. This translates into the condition
\begin{equation}
\frac{-\bar{Q}^2}{2 m_a\left(m_a-\sqrt{Q^2}\right)} < x_0 <1 , \label{eq:x0Cond}
\end{equation}
if $0< \sqrt{Q^2}< m_a-m_j$. If the latter condition is not met, any value for $x_0$ between zero and one can be used. 
Since the singular behavior occurs for $x \to 1$, applying the splitting function $\mathbf{V}_{ij}^a$ only for $x_0\leq x\leq 1$ still cancels the divergences. In addition, the independence of $\Delta\sigma^{\text{NLO}}$ on the choice of $x_0$ serves as a non-trivial check for the correct implementation of the subtraction procedure.
The integration of the dipole splitting function over the one-particle phase space 
\begin{equation}
\int \left[\dd p_i\left(Q^2,x,z_i\right)\right] =  \frac{1}{\left(4\pi\right)^{2-\varepsilon}}\frac{\left(P^2\right)^{-\varepsilon } }{\Gamma (1-\varepsilon )} \frac{-\bar{Q}^2}{x^2} \left(\frac{R(x)  \sqrt{\lambda_{aj}}}{-\bar{Q}^2}\right)^{2 \varepsilon -1} \int_{z_{-}}^{z_{+}}\dd{z_i} \left[\left(z_i-z_{-}\right)
   \left(z_{+}-z_i\right)\right]^{-\varepsilon}  \label{eq:FEIS_dip_PS}
\end{equation}
with the integration limits
\begin{equation}
z_{\pm}=\frac{1-x}{2} \frac{-\bar{Q}^2 \pm\sqrt{\lambda_{aj}}R(x)}{x m_j^2 -\bar{Q}^2 (1 - x)} \label{eq:int_limits_z}
\end{equation}
yields the singular behavior parameterized by $D=4-2\varepsilon$ dimensions. In the massless case $m_j=0$ the integration limits related to $z_i$ simplify to
\begin{equation}
z_{\pm}=\frac{1}{2}\left(1\pm R(x)\right).
\end{equation}

\subsection{The dipole splitting functions}
The functions $\mathbf{V}_{ij}^a$ in \cref{eq:dipole_FEIS} are provided for the four (SUSY)-QCD splitting processes
\begin{itemize}
 \item $q\to g(p_i)+q(p_j) \ $: $m_i=0$ and $m_{ij}=m_j=m_q$
\item $\tilde{q}\to g(p_i)+\tilde{q}(p_j) \ $: $m_i=0$ and $m_{ij}=m_j=m_{\tilde{q}}$
\item $g\to g(p_i)+g(p_j) \ $: $m_{ij}=m_i=m_j=0$
\item $g\to q(p_i)+\bar{q}(p_j) \ $: $m_i=m_j=m_{ij}=m_q=0$ .
\end{itemize}
The processes where the particles are exchanged through their corresponding antiparticles are formally identical to those given here and are therefore not listed separately.
The dipole splitting functions read explicitly
\begin{align}
\langle s| \mathbf{V}_{g q}^{a}|s'\rangle  &=8 \pi \alpha_s  \mu^{2\varepsilon} C_F \left(\frac{2}{2 - x_{ij,a} - z_j}-  2 + \frac{1}{2} h_g^{\textsc{RS}} z_i - \frac{m_j^2}{p_i\cdot p_j}\right)  \delta_{s s'}=\langle  \mathbf{V}_{g q}^{a}\rangle \delta_{s s'} \label{eq:FEIS_VgQ}\\
\langle s| \mathbf{V}_{g \tilde{q}}^{a}|s'\rangle  &=8 \pi \alpha_s  \mu^{2\varepsilon} C_F \left(\frac{2}{2 - x_{ij,a} - z_j}-  2 - \frac{m_j^2}{p_i\cdot p_j}\right)  \delta_{s s'}=\langle  \mathbf{V}_{g \tilde{q}}^{a}\rangle \delta_{s s'} \label{eq:FEIS_VgSq}\\
 \langle\mu|\mathbf{V}_{g g}^{a}|\nu\rangle &=16 \pi \alpha_s\mu^{2\varepsilon} C_A \left[-g^{\mu\nu} \left(\frac{1}{1+z_{i}-x_{ij,a}}+\frac{1}{2-z_{i}-x_{ij,a}}-2\right) + 
  \frac{h_g^{\textsc{RS}}}{2 p_i\cdot p_j} \mathcal{C}^{\mu\nu} \right]  \label{eq:FEIS_Vgg} \\
\langle\mu|\mathbf{V}_{q \overline{q}}^{a}|\nu\rangle &= 8 \pi \alpha_s \mu^{2\varepsilon}  T_F \left(-g^{\mu\nu} - \frac{2}{p_i\cdot p_j} \mathcal{C}^{\mu\nu}\right) .
\end{align}
In contrast to the work done in \cite{Catani:2002} and \cite{Kotko:2012agp}, the dipole splitting functions for processes involving quarks and gluons in \cref{eq:FEIS_VgQ} and \cref{eq:FEIS_Vgg} also include the number of helicity states of the gluon $h^{RS}_g$ in order to distinguish directly between the different variants of dimensional regularization schemes. The dipole splitting function for the squarks in \cref{eq:FEIS_VgSq} was derived by using the Eikonal approximation for a process involving the emission of a gluon off a squark.
The function in \cref{eq:FEIS_VgQ} for the splitting $q\to g + q$ is valid for a massive as well as a massless quark and the one in \cref{eq:FEIS_VgSq} for the splitting $\tilde{q} \to g \tilde{q}$ can also be applied to the process  $\tilde{g}\to g \tilde{g}$ since \cref{eq:FEIS_VgSq} only contains the soft limit.
The spin correlation tensor 
\begin{equation}
\mathcal{C}^{\mu\nu}=\left(z_i^{(m)} p_i^{\mu} - z_j^{(m)} p_j^\mu\right) \left(z_i^{(m)} p_i^{\nu} - z_j^{(m)} p_j^\nu\right)
\end{equation}
depending on the new variables 
\begin{equation}
  z_i^{(m)} = z_i - z_- = z_i-\frac{1}{2}\left(1-R(x)\right), \hspace{1cm}  z_j^{(m)} = z_j - z_- = z_j-\frac{1}{2}\left(1-R(x)\right)
\end{equation}
is constructed such that it reduces to $k_\perp^\mu k_\perp^\nu$ in the collinear limit as dictated by \cref{eq:altarelli-parisi} and is at the same time orthogonal to the direction of the emitter 
\begin{equation}
\tilde{p}_{ij}^\mu \mathcal{C}_{\mu\nu}=\tilde{p}_{ij}^\nu \mathcal{C}_{\mu\nu}=0. \label{eq:transversality_pij}
\end{equation}
The orthogonality allows to simplify the integration of the non-diagonal dipole functions in helicity space over the one-particle phase space in \cref{eq:FEIS_dip_PS} which is complicated due to the additional azimuthal correlations. However, the integral over the spin correlation tensor takes by Lorentz invariance (it can only depend on $\tilde{p}_{ij}$ and $\tilde{p}_a$) the form
\begin{equation} 
\int \left[\dd p_i\left(Q^2,P^2,z_i\right)\right] \mathcal{C}^{\mu\nu} =  - A_1 g^{\mu\nu}+ A_2 \frac{\tilde{p}_{ij}^\mu \tilde{p}_a^\nu+\tilde{p}_{ij}^\nu \tilde{p}_a^\mu}{\tilde{p}_{ij}\cdot \tilde{p}_a }- A_3  \frac{m_a^2 \tilde{p}_{ij}^\mu \tilde{p}_{ij}^\nu}{\left(\tilde{p}_{ij}\cdot \tilde{p}_a\right)^2} + A_4 \frac{\tilde{p}_a^\mu \tilde{p}^\nu_a}{m_a^2}\label{eq:factorizationTensorSplitting}.
\end{equation} 
At this point, note that the metric tensor multiplying $A_1$ is quasi-$D$-dimensional as momenta are kept in $D$ dimensions in all dimensional schemes.  Due to the transversality condition on $\mathcal{C}_{\mu\nu}$ in \cref{eq:transversality_pij} the term $A_4$ is zero and one finds additionally $A_1 = A_2$ such that the right hand side reduces to
\begin{equation}
     - A_1 \left(g^{\mu\nu}-  \frac{\tilde{p}_{ij}^\mu \tilde{p}_a^\nu+\tilde{p}_{ij}^\nu \tilde{p}_a^\mu}{\tilde{p}_{ij}\cdot \tilde{p}_a }\right)- A_3  \frac{m_a^2 \tilde{p}_{ij}^\mu \tilde{p}_{ij}^\nu}{\left(\tilde{p}_{ij}\cdot \tilde{p}_a\right)^2}.
\end{equation}
Therefore, $A_1$ can be disentangled by performing the azimuthal average over the transverse polarizations of the emitter 
\begin{equation}
A_1=\int \left[\dd p_i\left(Q^2,P^2,z_i\right)\right] \frac{1}{D-2} d_{\mu\nu}\left(\tilde{p}_{ij},\tilde{p}_a\right) \mathcal{C}^{\mu\nu}
\label{eq:averageA1}
\end{equation}
with the help of the polarization tensor 
\begin{equation}
 d^{\mu\nu}\left(\tilde{p}_{ij},\tilde{p}_a\right)= - g^{\mu\nu}+ \frac{\tilde{p}_{ij}^\mu \tilde{p}_a^\nu+\tilde{p}_{ij}^\nu \tilde{p}_a^\mu}{\tilde{p}_{ij}\cdot \tilde{p}_a }- m_a^2 \frac{ \tilde{p}_{ij}^\mu \tilde{p}_{ij}^\nu}{(\tilde{p}_{ij}\cdot \tilde{p}_a)^2} 
\end{equation}
which fulfills in $D$ dimensions  $d^{\mu\nu} d_{\mu\nu}=D-2$. The coefficient $A_3$ drops out in this computations since  $d_{\mu\nu}\left(\tilde{p}_{ij},\tilde{p}_a\right)\tilde{p}_{ij}^\mu \tilde{p}_{ij}^\nu =0$. Furthermore, $A_3$ is irrelevant because of the Slavnov-Taylor identity $\tilde{p}^\mu_{ij}\mathcal{M}_\mu=0$ which holds for any matrix element $\mathcal{M}_\mu$ in a Becchi-Rouet-Stora-Tyutin (BRST)-invariant theory where the polarization vector $\epsilon^\mu_\lambda(\tilde{p}_{ij})$ has been amputated \emph{if all other polarization vectors in $\mathcal{M}_\mu$ are transverse}. This means in particular that the spin-averaged splitting functions $\langle\mathbf{V}_{ij}^{a}\rangle$ emerging from \cref{eq:averageA1} are diagonal in helicity space, i.e proportional to $-g^{\mu\nu}$. Concretely, they are
\begin{align}
\langle\mathbf{V}_{g g}^{a}\rangle &=16 \pi \alpha_s \mu^{2\varepsilon} C_A \left[\frac{1}{1+z_{i}-x_{ij,a}}+\frac{1}{2-z_{i}-x_{ij,a}}-2+\frac{h_g^{\textsc{RS}}}{2(1-\varepsilon)}\left(z_+-z_{i}\right)(z_i-z_-)\right] \\
\langle\mathbf{V}_{q \overline{q}}^{a}\rangle &=8 \pi \alpha_s \mu^{2\varepsilon} T_F \left(1-\frac{2}{1-\varepsilon} (z_+ - z_i)(z_i-z_-)\right)
\end{align}
where $z_\pm$ correspond to the integration limits in \cref{eq:int_limits_z}.

\subsection{The integrated dipole functions} \label{sec:int_dipole_FEIS}
The integral of the spin-averaged dipole function $\langle\mathbf{V}^{a}_{ij}\rangle$ over the dipole phase space is defined as 
\begin{align}
\frac{\alpha_s}{2\pi}\frac{1}{\Gamma(1-\varepsilon)} \left(\frac{4\pi\mu^2}{-\bar{Q}^2}\right)^\varepsilon I^{a}_{ij} \left(x;\varepsilon\right)=\int \left[\dd p_i\left(Q^2,x,z_i\right)\right] \frac{1}{2 p_i\cdot p_j} \frac{1}{x} \langle\mathbf{V}^{a}_{ij}\rangle \label{eq:intIaijFEIS}
\end{align}
where $I^{a}_{ij}$ depends on the auxiliary variable $x$ (and $Q^2$). The cases $m_j\neq 0$ and $m_j=0$ for the process $q\to g q$ have to be treated separately due to different kinds of singular behavior. For this reason, the associated integrated dipole is marked with a hat $\hat{I}$ for $m_j=0$ to distinguish it from the massive case. By writing the dipole phase space in \cref{eq:FEIS_dip_PS} in the form 
\begin{equation}
\int_{z_{-}}^{z_{+}}\dd{z_i}  \left[\left(z_i-z_{-}\right) \left(z_{+}-z_i\right)\right]^{-\varepsilon} = (z_+ - z_-)^{1-2\varepsilon} \int_0^1 \dd{t} \left[(1-t) t\right]^{-\varepsilon}
\end{equation}
through the substitution $t=\frac{z_i - z_-}{z_+ - z_-}$, the integration of the splitting function becomes straightforward in terms of the Euler beta function 
\begin{align}
\beta(a,b)=\int_0^1 \dd{t} (1-t)^{a-1} t^{b-1}=\frac{\Gamma(a) \Gamma(b)}{\Gamma(a+b)}
\end{align}
and the Gaussian hypergeometric function 
\begin{equation}
\, _2F_1(a,b;c;z)=\frac{\Gamma(c)}{\Gamma(b) \Gamma(c-b)}\int_0^1 \dd{t} \frac{t^{b-1}(1-t)^{c-b-1}}{(1-z t)^a}.
\end{equation} 
The integrated counterparts $I^{a}_{ij}$ of \cref{eq:intIaijFEIS} read
\begin{multline}
I_{g q}^{a}\left(x;\varepsilon\right)=   \frac{2C_F}{v R(x) x^2} \frac{1}{(1-x)^{1+2 \varepsilon}} \left( \eta_j x+  (1-x)\right)^{2\varepsilon } \left(\frac{-\bar{Q}^{2}}{P^2}\right)^\varepsilon\left[ \left(\frac{\sqrt{\lambda_{aj}}R(x)}{\bar{Q}^2} + \frac{1}{4} h^{\textsc{RS}}_g z_- (z_+ -  z_-)\right) \beta(1-\varepsilon ,1-\varepsilon ) \right. \\ \left. +\frac{1}{4} h^{\textsc{RS}}_g (z_+ -  z_-)^2 \beta(1-\varepsilon ,2-\varepsilon )-I_1(-A(x);\varepsilon)\right] \label{eq:I_gqq_splitting}
\end{multline}
\begin{multline}
    \hat{I}_{g q}^{a}\left(x;\varepsilon\right)=\frac{2 C_F }{x^{2-\varepsilon}}\frac{1}{(1-x)^{1+\varepsilon}}
  \left[\left(\frac{1}{8} h^{\textsc{RS}}_g (1-R(x))-1\right) \beta(1-\varepsilon ,1-\varepsilon ) \right. \\ \left. +\frac{1}{4} h^{\textsc{RS}}_g R(x) \beta(1-\varepsilon ,2-\varepsilon )-\frac{1}{R(x)}I_1(-A(x);\varepsilon)\right] \label{eq:I_gqq_splitting_massless}
\end{multline}
\begin{multline}
I_{g \tilde{q}}^{a}\left(x;\varepsilon\right)=   \frac{2C_F}{v R(x) x^2} \frac{1}{(1-x)^{1+2 \varepsilon}} \left( \eta_j x+  (1-x)\right)^{2\varepsilon } \left(\frac{-\bar{Q}^{2}}{P^2}\right)^\varepsilon\left[ \left(\frac{\sqrt{\lambda_{aj}}R(x)}{\bar{Q}^2}\right) \beta(1-\varepsilon ,1-\varepsilon ) -I_1(-A(x);\varepsilon)\right] 
\end{multline}
\begin{multline}
I^a_{gg}\left(x;\varepsilon\right)=-\frac{2 C_A  }{R(x)x^{2-\varepsilon} }\frac{1}{(1-x)^{1+\epsilon}} \left[I_1\left(-A(x);\varepsilon\right)-I_1(\tilde{A}(x);\varepsilon)\right. \\ \left. -\frac{h_g^{\textsc{RS}}}{2(1-\varepsilon)} R(x)^3\beta\left(2-\varepsilon ,2-\varepsilon\right)+2 R(x) \beta\left(1-\varepsilon ,1-\varepsilon \right)\right] \label{eq:I_gg_splitting}
\end{multline}
\begin{align}
I^a_{q\bar{q}}\left(x;\varepsilon\right)=\frac{T_F  }{ x^{2-\varepsilon}} \frac{1}{(1-x)^{1+\epsilon}} \left(\beta(1-\varepsilon ,1-\varepsilon )-\frac{2}{1-\varepsilon }R(x)^2 \beta(2-\varepsilon ,2-\varepsilon )\right)
\end{align}
where the arguments $A(x)$ and $\tilde{A}(x)$ of the function
\begin{align}
I_1(z;\varepsilon)=&z \int_0^1 \dd{t} \frac{((1-t) t)^{-\varepsilon }}{1-z t}=z \beta(1-\varepsilon,1-\varepsilon) \, _2F_1(1,1-\varepsilon;2-2 \varepsilon ;z) \nonumber \\
=&-\ln (1-z)+\varepsilon\left(2 \operatorname{Li}_2(z)+\frac{1}{2} \ln ^2(1-z)\right)+\order{\varepsilon^2}
\label{eq:integral_I1}
\end{align}
are defined as
\begin{align}
A(x)=\frac{z_+ - z_-}{1 - x + z_-}, \hspace{1cm}  \tilde{A}(x)= \frac{z_+-z_-}{2-x-z_-}
\end{align}
and $v$ was defined in \cref{eq:v_and_eta}.
The expansion of $I_1(z;\varepsilon)$ in $\varepsilon$ is derived in \cref{sec:dipole_integrals} and is valid as long as its argument $z$ remains bounded between one and negative infinity $-\infty < z <1$ which is always fulfilled for $m_j\neq 0$. 
In the massless case $m_j=0$, the variables $A$ and $\tilde{A}$ take a similar form
\begin{align}
A(x)=\frac{2 \sqrt{1-w(x)^2}}{1-\sqrt{1-w(x)^2}}, \ \ \ \ \ \ \ \tilde{A}(x)=\frac{2 \sqrt{1-w(x)^2}}{1+\sqrt{1-w(x)^2}} \label{eq:ASimple}
\end{align}
when being expressed through the quantity
\begin{equation}
 w(x)=\frac{2 \sqrt{(1-x) (2-x(1-\eta_a))}}{3-2 x}
\end{equation}
where $\eta_a$ was defined in \cref{eq:v_and_eta}.
Writing $A(x)$ and $\tilde{A}(x)$ in this manner makes the relation
\begin{equation}
\frac{\tilde{A}(x)}{\tilde{A}(x)-1}=-A(x)
\end{equation}
apparent which allows to simplify the difference of the $I_1$ functions in \cref{eq:I_gg_splitting}
\begin{equation}
I_1\left(-A(x);\varepsilon\right)-I_1(\tilde{A}(x);\varepsilon)=2 I_1\left(-A(x);\varepsilon\right)
\end{equation}
by employing the identity 
\begin{equation}
I_1\left(z;\varepsilon\right)=-I_1\left(\frac{z}{z-1};\varepsilon\right) \label{eq:Pfaff_intI}
\end{equation}
which follows directly from the Pfaff transformation
\begin{align}
\, _2F_1\left(a,b;c;z\right)=(1-z)^{-a} \, _2F_1\left(a,c-b;c;\frac{z}{z-1}\right). \label{eq:Pfaff_trafo_2F1}
\end{align}
For $\varepsilon=0$ the functions $I_{ij}^{a}$ become singular at the endpoint $x\to 1$ giving the infrared divergence. Therefore, the integration over $x$ involving the tree matrix element squared can not be handled numerically yet. To allow for the extraction of the divergence in terms of $\varepsilon$ while being able to perform the mentioned integration numerically, the $[...]^+$-distribution defined as
\begin{equation}
g(x)=\left[g(x)\right]^{+}_{\left[a,b\right]}  + \delta(x-b) \int_{a}^b \dd{y} g(y) \label{eq:plus_distribution}
\end{equation}
provides a way around and serves as an artificially inserted zero to render the endpoint contribution finite. The endpoint part is then further decomposed into a finite $J_{ij}^{a;\text{NS}}$ and singular $J_{ij}^{a;\text{S}}$ piece where the latter contains the infrared poles such that we can write
\begin{align}
 & I_{gq}^{a}\left(x;\varepsilon\right) = C_F \left\{\left[ J_{gq}^{a}\left(x\right)\right]_+ + \delta(1-x) \left(J_{gq}^{a;\text{S}}\left(\varepsilon\right)+J_{gq}^{a;\text{NS}}\right) \right\}+\order{\varepsilon} \label{eq:I_gQa_splitting_distro} \\
  & I_{g\tilde{q}}^{a}\left(x;\varepsilon\right) = C_F \left\{\left[ J_{g\tilde{q}}^{a}\left(x\right)\right]_+ + \delta(1-x) \left(J_{g\tilde{q}}^{a;\text{S}}\left(\varepsilon\right)+J_{g\tilde{q}}^{a;\text{NS}}\right) \right\}+\order{\varepsilon} \label{eq:I_gSQa_splitting_distro} \\
    & I_{gg}^{a}\left(x;\varepsilon\right) = 2 C_A \left\{\left[ J_{gg}^{a}\left(x\right)\right]_+ + \delta(1-x) \left(J_{gg}^{a;\text{S}}\left(\varepsilon\right)+J_{gg}^{a;\text{NS}}\right)\right\}+\order{\varepsilon} \label{eq:I_gga_splitting_distro} \\
 & I_{q\bar{q}}^{a}\left(x;\varepsilon\right) = T_F \left\{\left[ J_{q\bar{q}}^{a}\left(x\right)\right]_+ + \delta(1-x) \left(J_{q\bar{q}}^{a;\text{S}}\left(\varepsilon\right)+J_{q\bar{q}}^{a;\text{NS}}\right)\right\}+\order{\varepsilon}. \label{eq:I_qqbara_splitting_distro} 
\end{align} 
The decomposition in \cref{eq:I_gQa_splitting_distro} holds similarly for the hatted and non-hatted versions. Note that the notation in \cref{eq:I_gQa_splitting_distro,eq:I_gSQa_splitting_distro,eq:I_gga_splitting_distro,eq:I_qqbara_splitting_distro} is purely symbolic: $\left[J_{ij}^{a}\left(x\right)\right]_+$ is not a "plus"-distribution itself but contains all the plus distributions.
The decomposition above is straightforward for the two cases involving either a soft or a collinear divergence through the identity
\begin{equation}
    \int_{x_0}^{1} \dd{x} \frac{1}{(1-x)^{1+\varepsilon}} f(x) = \left(-\frac{1}{\varepsilon}+\ln\left(1-x_0\right)\right) f\left(1\right) 
    + \int_{x_0}^{1} \dd{x} \frac{1}{1-x} \left(f(x)-f\left(1\right)\right) +\order{\varepsilon}.
\end{equation}
The gluon emission contributions for massive (s)quarks are 
\begin{align}
&\left[ J_{g q}^{a}\left(x\right)\right]_+ =  \frac{2}{x^2}\left[\frac{1}{1-x}\right]^+_{\left[x_0,1\right]}  \left(\frac{ (x-1)^2}{4(x(\eta_j - 1)+1)^2}-1 + \frac{1}{v R(x)}\ln(1+A(x))\right)   \label{eq:I_gqq_splitting_distro} \\
&\left[ J_{g \tilde{q}}^{a}\left(x\right)\right]_+ =  \frac{2}{x^2}\left[\frac{1}{1-x}\right]^+_{\left[x_0,1\right]}  \left( \frac{1}{v R(x)}\ln(1+A(x))-1\right)   \label{eq:I_gSQ_splitting_distro}
\end{align}
\begin{align}
&J_{g q}^{a;\text{S}}\left(\varepsilon\right)=J_{g \tilde{q}}^{a;\text{S}}\left(\varepsilon\right)=\frac{1}{\varepsilon}\left(1- \frac{1}{v}\ln(A+1)\right) \label{eq:J_QgQ1} \\
&J_{g q}^{a;\text{NS}}=J_{g \tilde{q}}^{a;\text{NS}}=\frac{2}{v}\left(\frac{1}{2}(v-\ln (A+1)) \ln\left(\frac{\eta_j}{(1-x_0)^2}\right)+v +\frac{1}{4}\ln^2(1+A)+\operatorname{Li}_2\left(-A\right) \right). \label{eq:J_QgQ2}
\end{align}
Note that $A$ is evaluated at $x=1$ in \cref{eq:J_QgQ1,eq:J_QgQ2} giving
\begin{equation}
A(1) = \frac{2 v}{2 \eta_j+1- v}. \label{eq:Aval1}
\end{equation} 
The continuum and endpoint contributions for the case of the splitting process $g\to q\bar{q}$ are
\begin{equation}
\left[J^a_{q\overline{q}}\left(x\right)\right]_{+}=  \left[\frac{1}{1-x}\right]^{+}_{\left[x_0,1\right]}\frac{1}{x^2} \left(1-\frac{1}{3}R(x)^2 \right)
\end{equation}
\begin{align}
J_{q\overline{q}}^{a;\text{S}}\left(\varepsilon\right)&=-\frac{2}{3 \varepsilon } \\
J_{q\overline{q}}^{a;\text{NS}}&=- \frac{10}{9}+\frac{2}{3} \ln\left(1-x_0\right).
\end{align}
Disentangling the infrared poles for massless quarks as well as gluons in the splittings $q\to g q$ and $g\to gg$ is more involved due to the fact that besides the factor $\frac{1}{\left(1-x\right)^{1+\varepsilon}}$ in \cref{eq:I_gqq_splitting_massless} and \cref{eq:I_gg_splitting} the function $I_1(-A;\varepsilon)$ diverges as well for $x\to 1$ which corresponds to a soft-collinear divergence. Since the expansion in $\varepsilon$ of $I_1(-A;\varepsilon)$ is not analytic for $x=1$, the hypergeometric function itself has to be placed inside the 
$[...]^+$-distribution which is achieved by introducing the argument of the hypergeometric function as new integration variable
\begin{equation}
y_A(x)= \frac{1}{A(x)} = (1-x) \mathcal{A}\left(x\right)
\end{equation}
which behaves analogously to $x$ in the singular region. This factorization is achieved by expanding both the numerator and the denominator of $A$ as given in \cref{eq:ASimple} with the term $[1+\sqrt{1-w(x)^2}]$ leading to 
\begin{equation}
\mathcal{A}\left(x\right)=\frac{2 ((1-\eta_a) x-2)}{ \rho (2 x-3-\rho)}
\end{equation}
with the abbreviation
\begin{equation}
\rho=\sqrt{1+4 \eta_a (x-1) x}. \label{eq:rho_var}
\end{equation}
In this new variable $y_A$ only the integral 
\begin{align}
\mathcal{I}_1\left(y_{0};\varepsilon\right)= \int_{0}^{y_{0}} \dd{y} \frac{1}{y^{1+\varepsilon}} I_1\left(-\frac{1}{y},\varepsilon\right)=-\frac{1}{2\varepsilon ^2}+\frac{\pi ^2}{12}-\operatorname{Li}_2\left(-\frac{1}{y_{0}}\right)+\order{\varepsilon}
\end{align}
has to be computed analytically which is outlined in \cref{sec:dipole_integrals}.
As we still want to perform the numerical integration of the $[...]^+$-distribution in terms of $x$, the derivative 
\begin{equation}
y_A'(x)=\pdv{y_A(x)}{x}=\frac{1}{\rho^3} \left( (3-4x) \eta_a -1\right)
\end{equation}
has to be included inside the "plus"-distribution.
For the special case $x=1$ it simply evaluates to
\begin{equation}
y_A'(1)=-\mathcal{A}(1)=-\eta_a -1.
\end{equation}
The explicit contributions to the decompositions in  \cref{eq:I_gQa_splitting_distro,eq:I_gga_splitting_distro} read then 
\begin{equation}
    \left[\hat{J}_{g q}^{a}\left(x\right)\right]_+ =  \frac{2}{x^2 }
  \left(-\frac{3}{4}\left[\frac{1}{1-x }\right]^{+}_{\left[x_0,1\right]}+ \left[y_A'(x) A(x) \ln\left(1+A(x) \right)\right]^{+}_{\left[x_0,1\right]}  \frac{\mathcal{A}(x)}{y_A'(x) R(x) } \right) 
\end{equation}
\begin{align}
    \hat{J}_{g q}^{a;\text{S}}\left(\varepsilon\right)=&\frac{1}{\varepsilon^2} +\frac{1}{\varepsilon} \left(\ln \left(1+\eta_a\right)+\frac{3}{2}\right)  \\
    \hat{J}_{g q}^{a;\text{NS}}=&  \frac{1}{2} \ln^2\left(1+\eta_a\right)-\frac{3}{2} \ln\left(1-x_0\right)+2 \operatorname{Li}_2\left(-A(x_0)\right)+\frac{7-r}{2}-\frac{\pi^2}{6}
\end{align}
\begin{equation}
 \left[ J^a_{gg} \left(x\right) \right]_{+}=  \frac{1}{x^2} \left(\left[y_A'(x) A(x) \ln\left(1+A(x) \right)\right]^{+}_{\left[x_0,1\right]} \frac{2\mathcal{A}(x)}{y_A'(x) R(x) } +\left[\frac{1}{1-x }\right]^{+}_{\left[x_0,1\right]}\left(\frac{R(x)^2}{6}-2\right) \right) 
\end{equation}
\begin{align}
J_{g g}^{a;\text{S}}\left(\varepsilon\right)&=\frac{1}{\varepsilon ^2}+\frac{1}{ \varepsilon}\left( \ln \left(1+\eta_a\right)+\frac{11}{6}\right) \label{eq:poles_gluon} \\
J_{g g}^{a;\text{NS}}&= \frac{1}{2}\ln ^2\left(1+\eta_a\right)-\frac{11}{6} \ln \left(1-x_0\right)+2 \operatorname{Li}_2\left(-A(x_0)\right)+\frac{67}{18}-\frac{\pi^2}{6} - \frac{r}{6}.
\end{align}


\subsection{Final expression}
\label{sec:FEIS-final}
We are now ready to present the explicit form of the insertion operator $\mathbf{I}$ defined in \cref{eq:defI} following the same notation used by Catani and Seymour in Ref. \cite{Catani:2002}. Therefore, we consider a process with one colored initial particle carrying the momentum $p_a$ and another not necessarily colored particle with momentum $p_b$. 
The final result for the auxiliary cross section can then be written as  
\begin{equation}
    \int_{m+1} \dd\sigma^{\text{A}}_a = \int_{x_0}^1 \dd{x}  \int_m \left[\dd\sigma^\text{B}_a\left(\tilde{p}_a(x)\right) \frac{\mathcal{F}_{\tilde{a}}}{\mathcal{F}_a} \otimes \mathbf{I}_{m,a}\left(x;\varepsilon\right)\right] \label{eq:dAFEIS}
\end{equation}
with the new insertion operator
\begin{equation}
    \mathbf{I}_{m,a}\left(x;\varepsilon,\mu^2;\{p_i,m_i\},p_a\right) = - \frac{\alpha_s}{2\pi} \frac{(4\pi)^\varepsilon}{\Gamma(1-\varepsilon)} \sum_j \mathbf{T}_j\cdot \mathbf{T}_{a} \left(\frac{\mu^2}{-\bar{Q}^2}\right)^\varepsilon \frac{1}{\mathbf{T}_j^2} \mathcal{V}_j\left(x;Q^2,m_j;\varepsilon\right).
\end{equation}
The factor $\nicefrac{\mathcal{F}_{\tilde{a}}}{\mathcal{F}_a}$ in \cref{eq:dAFEIS}  with
\begin{align}
    \mathcal{F}_a=2 \lambda^\frac{1}{2}((p_a+p_b)^2,m_a^2,m_b^2), \ \ \ \ \ \ \ \
    \mathcal{F}_{\tilde{a}}=2 \lambda^\frac{1}{2}((\tilde{p}_a(x)+p_b)^2,m_a^2,m_b^2) \label{eq:fluxes}
\end{align}
is responsible for the correct flux factor. The flavor functions $\mathcal{V}_j$
already incorporate the correct counting of the symmetry factors for the transition from $m+1$ particles to $m$ particles:
\begin{itemize}
    \item If $j$ is a massive quark (or antiquark), then
\begin{equation}
    \mathcal{V}_q(x,Q^2,m_q;\varepsilon) = I_{gq}^{a}\left(x;\varepsilon\right).
\end{equation}
\item If $j$ is a massless quark (or antiquark), then
\begin{equation}
    \mathcal{V}_q(x,Q^2,0;\varepsilon)=C_F \left[ J_{gq}^{a}\left(x\right)\right]_+  
     +\delta(1-x) \left\{C_F \left(\frac{1}{\varepsilon^2}+ \chi\left(Q^2\right)  -\frac{3}{2} \ln\left(1-x_0\right)\right) + \Gamma^{\textsc{RS}}_q(\varepsilon) + K_q \right\}
\end{equation}
with
\begin{equation}
    \chi\left(Q^2\right)=\frac{1}{\varepsilon} \ln \left(1+\eta_a\right)  + \frac{1}{2} \ln^2\left(1+\eta_a\right)+2 \operatorname{Li}_2\left(-A(x_0)\right).
\end{equation}
\item The flavor kernel for a gluon $j$ is
\begin{multline}
    \mathcal{V}_g(x,Q^2,0;\varepsilon)=C_A \left[ J_{gg}^{a}\left(x\right)\right]_+ + T_F N_f \left[ J_{Q\bar{Q}}^{a}\left(x\right)\right]_+ 
     + \delta(1-x) \left\{C_A \left(\frac{1}{\varepsilon^2}+ \chi\left(Q^2\right)-\frac{11}{6} \ln\left(1-x_0\right)  \right) \right. \\ \left.  + \frac{2}{3} T_F N_f  \ln\left(1-x_0\right) + \Gamma_g^{\textsc{RS}}(\varepsilon) + K_g \right\}.
\end{multline}
\end{itemize}
The functions $\Gamma_j$ for gluons and massless quarks (antiquarks) are
\begin{equation}
\Gamma_q^{\textsc{RS}}\left(\varepsilon\right)=\frac{1}{\varepsilon}  \gamma_q - \tilde{\gamma}^{\textsc{RS}}_q, \ \ \ \ \ \ \ \ \
\Gamma_g^{\textsc{RS}}(\varepsilon)= \frac{1}{\varepsilon} \gamma_g - \tilde{\gamma}^{\textsc{RS}}_g
\end{equation}
 with the flavor constants
\begin{equation}
\gamma_q = \frac{3}{2} C_F, \ \ \ \ \ \ \ \ \   
\gamma_g = \frac{11}{6} C_A- \frac{2}{3}  T_F N_f 
\end{equation}
and the regularization scheme dependent terms
 \begin{equation}
     \tilde{\gamma}^{\textsc{RS}}_g = \frac{r}{6} C_A, \ \ \ \ \  \ \ \ \  \tilde{\gamma}^{\textsc{RS}}_q = \frac{r}{2} C_F.
 \end{equation}
The constants $K_a$ are defined as
\begin{equation}
K_q =\left(\frac{7}{2}- \frac{\pi^2}{6}\right) C_F, \ \ \ \ \ \ \ \ \
K_g = \left(\frac{67}{18}- \frac{\pi^2}{6}\right) C_A -\frac{10}{9} T_F N_f.
\end{equation}

\section{Initial-state emitter and final-state spectator}
The dipole $\mathcal{D}^{ai}_j$ in \cref{eq:dipole_formula} is defined as
\begin{equation}
\mathcal{D}^{ai}_j=-\frac{1}{2 p_a\cdot p_i} \frac{1}{x_{ij,a}}\tensor*[_{m,\widetilde{ai}}]{\bra{\dots,\widetilde{j},\dots;\widetilde{ai},\dots}}{}\frac{\mathbf{T}_j\cdot \mathbf{T}_{ai}}{\mathbf{T}_{ai}^2} \mathbf{V}^{ai}_j \ket{\dots,\widetilde{j},\dots;\widetilde{ai},\dots}_{m,\widetilde{ai}} \label{eq:dipole_IEFS}
\end{equation}
where $\mathbf{V}^{ai}_j$ describes the splitting process $a \to i + \widetilde{ai}$. The tree-level matrix element is obtained from the original matrix element with $(m+1)$-particles in the final state by replacing the momentum $p_a$ of the particle $a$ in the tree-level matrix element by the dipole momentum $\tilde{p}_{ai}$, the momentum $p_j$ of $j$ by $\tilde{p}_j$ and discarding the final-state particle $i$. 
Similar to the previous section, we consider only the case where the masses of $a$ and $\widetilde{ai}$ are identical.
\subsection{Kinematics and phase space factorization}
The case of an initial-state emitter and final-state spectator is kinematically identical to the case of a final-state emitter and an initial-state spectator after switching the roles played by $\widetilde{ij}$ and $a$. Particle $j$ takes over the role of the spectator and the associated dipole momenta are relabelled accordingly as $\tilde{p}_{ij}\to \tilde{p}_j$ and $\tilde{p}_{a}\to \tilde{p}_{ai}$. Therefore, the kinematics from \cref{sec:FEIS_kinematics} can be adopted completely. 

\subsection{The dipole splitting functions}
The function $\mathbf{V}^{ai}_j$ in \cref{eq:dipole_IEFS} for the SUSY-QCD splitting process 
\begin{itemize}
\item  $\tilde{q}(p_a)\to g(p_i) + \tilde{q} \ $: $m_i=0$ and $m_a=m_{\tilde{q}}$
\end{itemize}
in presence of a massive emitter $\widetilde{ai}$ reads
\begin{equation}
\langle\mathbf{V}^{\tilde{q} g}_{j}\rangle = 8 \pi \alpha_s C_F \mu^{2\varepsilon} \left(\frac{2}{2 - x_{ij,a} - z_j} - 2 - \frac{m_a^2  x_{ij,a}}{p_a\cdot p_i}\right). \label{eq:V_IE-FS}
\end{equation}
The same function holds for the gluino splitting process $\tilde{g} \to g + \tilde{g}$ 
as well as for $q \to g q$ involving a massive quark as \cref{eq:V_IE-FS} only accounts for the soft limit. In the case of the gluino, it is only necessary to replace the color factor $C_F$ in \cref{eq:V_IE-FS} by $C_A$.

\subsection{The integrated dipole functions}
We define the integral of the spin-averaged dipole function $\langle\mathbf{V}^{ai}_j\rangle$ over the dipole phase space as 
\begin{align}
\frac{\alpha_s}{2\pi}\frac{1}{\Gamma(1-\varepsilon)} \left(\frac{4\pi\mu^2}{-\bar{Q}^2}\right)^\varepsilon I^{a,\widetilde{ai}}_j \left(x;\varepsilon\right)=\int \left[\dd p_i\left(Q^2,x,z_i\right)\right] \frac{1}{2 p_a\cdot p_i} \frac{1}{x} \langle\mathbf{V}^{ai}_j\rangle . \label{eq:int_dipole_IEFS}
\end{align}
In our case, it is not necessary to differentiate between the number of polarizations $n_s(\widetilde{ai})$ ($n_s(a)$) of $\widetilde{ai}$ ($a$) in contrast to \cite{Catani:1996vz} as we always have the same number of polarizations of $a$ and $\widetilde{ai}$.
For the only splitting function considered in this section, the integral over the dipole phase space can be performed in a straightforward manner through a partial fraction decomposition and the application of the hypergeometric as well Euler's Beta function giving
\begin{multline}
I^{\tilde{q} \tilde{q}}_{j}\left(x;\varepsilon\right)=\frac{2 C_F}{v R(x) x^2}\frac{1}{(1-x)^{1+2 \varepsilon} } \left(\eta_j x+ (1-x)
   \right)^{2\varepsilon }\left(\frac{-\bar{Q}^2}{P^2}\right)^\varepsilon  \\
  \times\left(I_1(-A(x);\varepsilon)- x I_1(-B(x);\varepsilon)+2 \eta_a x^2\frac{ (\eta_j-1) x+ 1}{1 -  v R(x)} I_2(-B(x);\varepsilon) \right) \label{eq:int_IEFSmassive}
\end{multline}
\begin{equation}
\hat{I}^{\tilde{q} \tilde{q}}_{_j}\left(x;\varepsilon\right)=\frac{2 C_F }{R(x) x^{2-\varepsilon}} \frac{1}{\left(1-x\right)^{1+\varepsilon}}  \left( I_1\left(-A(x);\varepsilon\right)-x I_1\left(-B(x);\varepsilon\right) +\frac{x}{2} (R(x)+1) I_2\left(-B(x);\varepsilon\right)\right)\label{eq:int_IEFSmassless}
\end{equation}
where the hat separates again the cases $m_j\neq 0$ and $m_j=0$. The variable $B$ is defined as 
\begin{align}
B(x)=\frac{z_+ - z_-}{z_-}=\frac{-2 \sqrt{\lambda_{aj}}R(x)}{\bar{Q}^2+\sqrt{\lambda_{aj}}R(x)}
\end{align}
and can be written in the massless case as 
\begin{equation}
B(x)=\frac{2 \sqrt{1-u(x)^2}}{1-\sqrt{1-u(x)^2}} \label{eq:B(u)} \hspace{1cm} \text{with} \hspace{1cm} u(x)^2=4 (1-x)x\eta_a.
\end{equation}
The variable $A(x)$ as well as the function $I_1(z;\epsilon)$ were introduced in \cref{sec:int_dipole_FEIS}. The function
\begin{align}
I_2(z;\varepsilon)&=z \int_0^1 \dd{t} \frac{((1-t) t)^{-\varepsilon }}{(1-z t)^2}=z \beta(1-\varepsilon,1-\varepsilon) \, _2F_1(2,1-\varepsilon;2-2 \varepsilon ;z) \nonumber  \\
&=\frac{z}{1-z}+\varepsilon \frac{2-z}{z-1}\ln(1-z)+\order{\varepsilon^2} \label{eq:I2_func}
\end{align}
is defined similarly to $I_1(z;\epsilon)$ but with a different argument set of the hypergeometric function. The extraction of the divergences proceeds again through the application of "plus"-distribution 
\begin{equation}
 I^{\tilde{q} \tilde{q}}_{j}\left(x;\varepsilon\right)=C_F \left\{\left[J^{\tilde{q} \tilde{q}}_{j}\left(x\right)\right]_+ +\delta\left(1-x\right) \left(J_{j}^{\tilde{q} \tilde{q};\text{S}}\left(\varepsilon\right)+J_{j}^{\tilde{q} \tilde{q};\text{NS}}\right)\right\}+\order{\varepsilon} \label{eq:decomposition_IEFS}.
\end{equation}
Performing this decomposition in the massless case is as peculiar as in the case of the gluon splitting function. However, it is possible to proceed in the same way. The divergent pieces given by $I_1\left(-A(x);\varepsilon\right)$ and $I_1\left(-B(x);\varepsilon\right)$ in \cref{eq:int_IEFSmassless} lead to the same integral $\mathcal{I}_1$ that already appeared in \cref{sec:int_dipole_FEIS}. The poles that arise through $I_2\left(-B;\varepsilon\right)$ can now be disentangled in a very similar manner by introducing the variable 
\begin{equation}
y_B(x)=\frac{1}{B(x)}=(1-x) \mathcal{B}\left(x\right)
\end{equation}
as new integration variable. This factorization is achieved as in the case of $A(x)$ by expanding $B(x)$ written as in \cref{eq:B(u)} with $1+\sqrt{1-u^2}$ which yields
\begin{equation}
\mathcal{B}\left(x\right)=\frac{2 \eta_a x}{\rho ^2+\rho }
\end{equation}
where $\rho$ was defined in \cref{eq:rho_var}. For the integration of the function $I_2\left(-B;\varepsilon\right)$ the integral  
\begin{equation}
\mathcal{I}_2(y_0;\varepsilon)=\int_0^{y_0} \dd{y}\frac{1}{y^{1+\varepsilon}} I_2\left(-\frac{1}{y};\varepsilon\right)=\frac{1}{2 \varepsilon }+\ln\left(1+\frac{1}{y_0}\right)+\order{\varepsilon}
\end{equation}
is calculated in \cref{sec:dipole_integrals}.
Connected to the transition from $x$ to $y_B$ the derivative 
\begin{equation}
y_B'(x)=\pdv{y_B(x)}{x}=\frac{\eta_a}{\rho^3}  (1-2 x)
\end{equation}
has to be placed inside the $[\dots]^+$-distribution. It simplifies to 
\begin{equation}
y_B'(1)=-\mathcal{B}(1)=-\eta_a
\end{equation}
for $x=1$. 
With the knowledge of the integrals $I_1$, $I_2$, $\mathcal{I}_1$ and $\mathcal{I}_2$ we can give the different contributions to \cref{eq:decomposition_IEFS} which read for the massive case 
\begin{equation}
\left[J^{\tilde{q} \tilde{q}}_{j}\left(x\right)\right]_+ =  \left[\frac{1}{1-x}\right]^+_{\left[x_0,1\right]} \frac{2}{v R(x)x^2} \left(x \ln(1+B(x)) -\ln(1+A(x)) - 4 \eta_a x^2 \frac{(\eta_j-1)x +1}{1 - v^2 R(x)^2} v R(x)  \right)
\end{equation}
\begin{align}
J_{j}^{\tilde{q} \tilde{q};\text{S}}\left(\varepsilon\right)=&\frac{1 }{\varepsilon}\left(1 - \frac{1}{v}\ln\left(  \frac{1+B}{1+A} \right)\right) \label{eq:JIEFSS}\\
J_{j}^{\tilde{q} \tilde{q};\text{NS}}=&\frac{1}{v}\left(\left(v+\ln \left(\frac{1+A}{1+B}\right)\right) \ln \left(\frac{\eta_j}{(1-x_0)^2}\right)+ \ln(1+B) \nonumber  \right. \\
&\left. +\frac{1}{2} \left[\ln^2(1+B)-\ln^2(1+A)\right]+2\operatorname{Li}_2\left(-B\right) -2\operatorname{Li}_2\left(-A\right)\right) \label{eq:JIEFSNS}
\end{align}
and for the massless case
\begin{multline}
\left[\hat{J}^{\tilde{q} \tilde{q}}_{j}\left(x\right)\right]_+ = \frac{2}{R(x) x} \left(\left[y_B'(x) B(x)\ln\left(1+B(x)\right)\right]^{+}_{\left[x_0,1\right]} \frac{\mathcal{B}(x)}{y_B'(x)} \right.\\
\left. -\left[y_A'(x) A(x) \ln\left(1+A(x)\right) \right]^{+}_{\left[x_0,1\right]}   \frac{\mathcal{A}(x)}{x y_A'(x)    }-\left[y_B'(x) \frac{B^2(x)}{1+B(x)}\right]^{+}_{\left[x_0,1\right]}  \frac{\mathcal{B}(x)}{2 y_B'(x)} \left(1+R(x)\right) \right) 
\end{multline}
\begin{align}
\hat{J}^{\tilde{q} \tilde{q};\text{S}}_{j}\left(\varepsilon\right)&=\frac{1}{\varepsilon}\left(1-\ln \left(\frac{1+\eta_a}{\eta_a}\right)\right) \\
\hat{J}_{j}^{\tilde{q} \tilde{q};\text{NS}}&=\frac{1}{2}\ln ^2\left(\eta_a\right)+\ln \left(\eta_a\right)-\frac{1}{2} \ln ^2\left(1+\eta_a\right)  -2\operatorname{Li}_2\left(-A(x_0)\right)+2\operatorname{Li}_2\left(- B(x_0)\right)+2\ln
   \left(1+ B(x_0)\right).
\end{align}
Note that $A$ and $B$ in \cref{eq:JIEFSS,eq:JIEFSS} are evaluated at $x=1$, i.e.
\begin{equation}
    B(1)=\frac{2 v}{1-v}
\end{equation}
where $v$ was defined in \cref{eq:v_and_eta}. The corresponding expression for $A$ is given in \cref{eq:Aval1}.

\subsection{Final expression}
By using the same labels as in \cref{sec:FEIS-final}, the auxiliary cross section for initial-state singularities with final-state spectators can be written as
    \begin{equation}
    \int_{m+1} \dd\sigma^{\text{A}}_a = \sum_{a'} \int_{x_0}^1 \dd{x}  \int_m \left[\dd\sigma^\text{B}_{a'}\left(\tilde{p}_{a}(x)\right) \frac{\mathcal{F}_{\tilde{a}}}{\mathcal{F}_a} \otimes \mathbf{I}_{m,aa'}\left(x;\varepsilon\right)\right] \label{eq:dAI_IEFS}
\end{equation}
with the insertion operator
\begin{equation}
    \mathbf{I}_{m,aa'}\left(x;\varepsilon,\mu^2;\{p_i,m_i\},p_a\right) = - \frac{\alpha_s}{2\pi} \frac{(4\pi)^\varepsilon}{\Gamma(1-\varepsilon)} \sum_j \mathbf{T}_j\cdot \mathbf{T}_{a'} \left(\frac{\mu^2}{-\bar{Q}^2}\right)^\varepsilon \frac{1}{\mathbf{T}_{a'}^2} \mathcal{V}^{a,a'}\left(x;Q^2,m_j;\varepsilon\right).
\end{equation}
The flavour functions $\mathcal{V}^{a, a'}$ are related to the integrated dipoles defined in \cref{eq:int_dipole_IEFS} via
\begin{equation}
    \mathcal{V}^{a, a'}(x,Q^2,m_j;\varepsilon) =  I^{a a'}_{j}\left(x;\varepsilon\right).
\end{equation}

\section{Initial-state emitter and initial-state spectator}
The dipole for emitter and spectator both from the initial state is defined as
\begin{equation}
\mathcal{D}^{ai,b}=\frac{1}{-2 p_a\cdot p_i} \frac{1}{x_{i,ab}} \tensor*[_{m,ab}]{\bra{\tilde{1},...,\widetilde{m+1};\widetilde{ai},b}}{}\frac{\mathbf{T}_b\cdot \mathbf{T}_{ai}}{\mathbf{T}_{ai}^2} \mathbf{V}^{ai,b}\ket{\tilde{1},...,\widetilde{m+1};\widetilde{ai},b}_{m,ab} \label{eq:dipole_IEIS}
\end{equation}
where the $m$-particle matrix element is obtained by discarding the particle $i$ in the $(m+1)$-particle matrix element and rescaling the momenta $p_k$ of \emph{all} other final state particles to their dipole analogues $\tilde{p}_k$ as well as $p_a$ to $\tilde{p}_{ai}$  while the momentum of the spectator $p_b$ remains \emph{unchanged}. The operator $\mathbf{V}^{ai,b}$ in \cref{eq:dipole_IEIS} describes the splitting $a \to \widetilde{ai} + i$.

\subsection{Kinematics and phase space factorization}

\begin{figure}[h]
    \centering

    \includegraphics{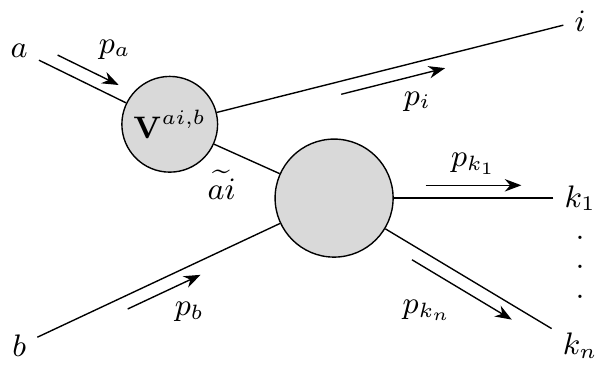}
    \caption{Diagrammatic interpretation of the dipole $\mathcal{D}^{ai,b}$ and the associated splitting function $\mathbf{V}^{ai,b}$.}
    \label{fig:3legsp}
\end{figure}

\begin{figure}
    \centering
\includegraphics{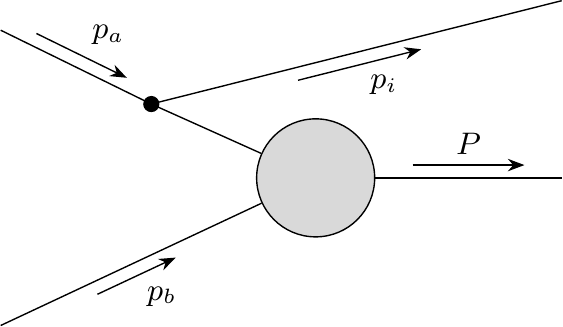}
\hspace{2cm}
    \includegraphics{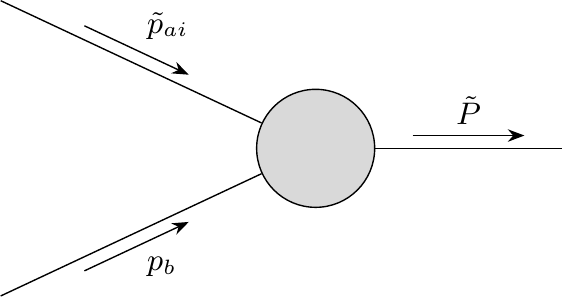}
    \caption{Kinematics for an initial-state emitter and an initial-state spectator in the original momenta (left) and the dipole momenta (right).}
    \label{fig:IEIS-dipoleKinematics}
\end{figure}

For the parametrization of the divergences we introduce the auxiliary variables
\begin{align}
x_{i,ab}=\frac{p_a\cdot p_b-p_i\cdot p_a-p_i\cdot p_b}{p_a\cdot p_b}, \ \ \ \ \ \ \
y = \frac{p_a\cdot p_i}{p_a\cdot p_b}
\end{align}
which behave in the soft limit $p_i^\mu \to 0$ as
\begin{equation}
x_{i,ab} \to 1, \ \ \ \ \ \ \ \ \ \  y\to 0.
\end{equation}
The sum of all outgoing momenta $p_k$ except for the soft gluon is denoted by
\begin{equation}
P = p_a + p_b -p_i=\sum_k p_k,
\end{equation}
cf.\ Figs.\ \ref{fig:3legsp} and \ref{fig:IEIS-dipoleKinematics}. Furthermore, it is convenient to define the abbreviations
\begin{align}
\lambda_{ab} &= \lambda\left(s,m_a^2,m_b^2\right)=\overline{s}^2 - 4 m_a^2 m_b^2 \\
\overline{s} &= s -m_a^2-m_b^2.
\end{align}
The construction of the dipole momenta is different from the previous two cases. Instead of modifying only the momenta of emitter and spectator, the momentum of the spectator $p_b$ remains unchanged whereas all other momenta are modified. The new momenta 
\begin{align}
\tilde{p}_{ai}^\mu &= \sqrt{\frac{\lambda\left(P^2,m_a^2,m_b^2\right)}{\lambda_{ab}}} p_a^\mu +\left(\frac{P^2-m_a^2-m_b^2}{2 m_b^2}-\frac{p_a\cdot p_b}{m_b^2}\sqrt{\frac{\lambda\left(P^2,m_a^2,m_b^2\right)}{\lambda_{ab}}}\right) p_b^\mu\\
\tilde{P}^\mu &= \tilde{p}_{ai}^\mu + p_b^\mu \label{eq:PtildeIEIS}
\end{align}
are then built from the requirement to retain the mass-shell relations $\tilde{p}_{ai}^2=m_{ai}^2$ and $\tilde{P}^2 = P^2$. The outgoing momenta $p_k$ except for $p_i$ are modified by a Lorentz transformation
\begin{equation}
\tilde{p}_k^\mu = \tensor{\Lambda}{^\mu_\nu} p_k^\nu
\end{equation}
with
\begin{equation}
\tensor{\Lambda}{^\mu_\nu} = \tensor{g}{^\mu_\nu} - \frac{\left(P+\tilde{P}\right)^\mu\left(P+\tilde{P}\right)_\nu}{P^2+ P\cdot\tilde{P}} + \frac{2 \tilde{P}^\mu P_{\nu}}{P^2}.
\end{equation}
If follows from direct calculation that $\tensor{\Lambda}{^\mu_\nu}$ indeed leaves the Minkowski metric invariant $\tensor{\Lambda}{_\rho^\mu}\tensor{\Lambda}{^\rho^\nu} = \tensor{g}{^\mu^\nu}$ such that it can be verified easily that the new momenta $\tilde{p}_k$ obey the on-shell condition $\tilde{p}_k^2=m_k^2$. The definition of these momenta coincides with \cite{Dittmaier:1999mb}. 
In order to ensure that $\lambda\left(P^2,m_a^2,m_b^2\right)$ remains positive, so that the dipole momenta take only real values, the kinematical lower bound
\begin{equation}
x_{i,ab} > x_0\geq \hat{x} = \frac{2 m_a m_b}{\overline{s}} \label{eq:xhatIEIS}
\end{equation}
on $x_{i,ab}$ has to be enforced. For values of $x_{i,ab}$ below $x_0$ the splitting functions $\mathbf{V}^{ai,b}$ are set to zero. The dependence on the lower bound $x_0$ must cancel out and can therefore be chosen arbitrarily which offers the possibility to check whether the implementation of the subtraction procedure is correct. The factorization of the single-particle phase space $\left[\dd{p_i}(s,x,y)\right]$ from the $(m+1)$-particle phase space $\dd{\phi}_{m+1}\left(p_i,P;p_a+p_b\right)$ is derived in \cref{sec:PSIE-IS}. It corresponds to a convolution over $x$ which plays the role of $x_{i,ab}$
\begin{equation}
    \int \dd{\phi}_{m+1}\left(p_i,P;p_a+p_b\right) \theta(x_{i,ab}-x_0) = \int_{x_0}^1 \dd{x} \int\dd{\phi}_m\left(\tilde{p}_{k}(x);\tilde{p}_{ai}(x)+p_b\right) \int \left[\dd{p_i}(s,x,y)\right]. \label{eq:PSIEIS}
\end{equation}
In $D=4-2\varepsilon$ dimensions the dipole phase space becomes
\begin{equation}
\int \left[\dd{p_i}(s,x,y)\right] = \frac{\overline{s}^{2-2\varepsilon}}{(4\pi)^{2-\varepsilon}\Gamma(1-\varepsilon)} \frac{s^{-\varepsilon}}{\sqrt{\lambda_{ab}}^{1-2\varepsilon}} \int_{y_-}^{y_+} \dd{y} \left[(y-y_-)(y_+ - y)\right]^{-\varepsilon} 
\end{equation}
where the integration boundaries read
\begin{equation}
y_\pm = \frac{1-x}{2s} \left(\overline{s}+2m_a^2 \pm \sqrt{\lambda_{ab}}\right).
\end{equation}
From \cref{eq:PtildeIEIS} it can be deduced that the c.m. energy 
\begin{equation}
    \tilde{s} = P^2= \bar{s} x + m_a^2 +m_b^2 \label{eq:stildeIEIS}
\end{equation} 
of the reduced phase space $\dd{\phi}_m\left(\tilde{p}_{k}(x);\tilde{p}_{ai}(x)+p_b\right)$ is already determined through $x$ and the original c.m. energy $\sqrt{s}$.

\subsection{The dipole splitting function}
The dipole function $\mathbf{V}^{ai,b}$ in \cref{eq:dipole_IEIS} for the SUSY-QCD process
\begin{itemize}
 \item $\tilde{q}(p_a)\to  g(p_i)+\tilde{q} \ $: $m_i=0$ and $m_{ai}=m_a=m_{\tilde{q}}$
\end{itemize}
reads 
\begin{equation}
\langle \mathbf{V}^{\tilde{q}g,b}\rangle = 8 \pi \alpha_s \mu^{2\varepsilon} C_F \left(\frac{2}{1-x_{i,ab}}-2-\frac{x_{i,ab} m_a^2}{p_a\cdot p_i}\right). \label{eq:splittingV-IE-IS}
\end{equation}
The dipole splitting functions for the processes involving a gluino $\tilde{g}\to g  \tilde{g} $ and a massive quark $q \to g q$ are for the pure soft limit identical to \cref{eq:splittingV-IE-IS} where only the color factor $C_F$ has to be replaced by $C_A$ for the gluino. For this reason, only the squark splitting function is treated in the following without losing generality. The same splitting function holds if the squark is replaced by an antisquark.

\subsection{The integrated dipole functions}
In complete analogy to previous cases, the integrated dipole for the case of emitter and spectator both from the initial state is defined as  
\begin{equation}
\frac{\alpha_s}{2\pi} \frac{1}{\Gamma(1-\varepsilon)} \left(\frac{4\pi\mu^2}{\bar{s}}\right)^{\varepsilon} I^{a,\widetilde{ai},b}(x;\varepsilon) = \int \left[\dd{p_i}(s,x,y)\right]  \frac{1}{2 p_a\cdot p_i} \frac{1}{x_{i,ab}} \langle\mathbf{V}^{ai,b}\rangle \label{eq:IIEIS}
\end{equation}
and the factorized phase space of the gluon can be turned into the convenient form
\begin{equation}
\int_{y_{-}}^{y_{+}}\dd{y}  \left[\left(y-y_{-}\right) \left(y_{+}-y\right)\right]^{-\varepsilon} = (y_+ - y_-)^{1-2\varepsilon} \int_0^1 \dd{t} \left[(1-t) t\right]^{-\varepsilon}
\end{equation}
via the substitution $t=\frac{y - y_-}{y_+ - y_-}$. 
By expressing the denominator in the dipole as $2 p_a\cdot p_i=y \overline{s} $ and with the help of the already known integrals $I_1(z;\varepsilon)$ and $I_2(z;\varepsilon)$ defined in \cref{eq:integral_I1} and \cref{eq:I2_func}, the integration of the splitting function in \cref{eq:splittingV-IE-IS} results in
\begin{align}
 I^{\tilde{q}\tilde{q},b}(x;\varepsilon)  = \frac{C_F}{\sqrt{\lambda_{ab}}} \frac{2}{ (1-x)^{1+2 \varepsilon}}\left(\frac{ s}{\bar{s}}\right)^{\varepsilon } \left(\frac{2 m_a^2 s}{d_1} I_2(-C;\varepsilon)-\bar{s}
   I_1(-C;\varepsilon)\right)
\end{align}
where the new auxiliary variables $C$ and $d_1$ are defined as
\begin{align}
C &= \frac{y_+ - y_-}{y_-} = \frac{2 \sqrt{\lambda_{ab}}}{d_1}\\
d_1&=\bar{s} + 2 m_a^2 - \sqrt{\lambda_{ab}}.
\end{align}
Since only massive initial states are considered, the argument $-C$ of the functions $I_1$ and $I_2$ does not diverge and we are allowed to use their associated expansions in $\varepsilon$. As explained in \cref{sec:int_dipole_FEIS}, the soft divergence can be disentangled with the help of the $[\dots]^+$-prescription
\begin{equation}
I^{\tilde{q}\tilde{q},b}(x;\varepsilon) = C_F\left\{\left[J^{\tilde{q}\tilde{q},b}(x)\right]_+ + \delta(1-x) \left[J^{\tilde{q}\tilde{q},b;\text{S}}(\varepsilon)+J^{\tilde{q}\tilde{q},b;\text{NS}}\right]\right\}+\order{\varepsilon}. \label{eq:IEISIcont}
\end{equation}
The continuum part of \cref{eq:IEISIcont} that contains the "plus"-distribution is given by
\begin{equation}
 \left[J^{\tilde{q}\tilde{q},b}(x)\right]_+ = 2 \left[\frac{1}{1-x}\right]^+_{[x_0,1]}(d_2 \ln(1+C)-1) \label{eq:IEIS_cont}.
\end{equation}
The endpoint parts are
\begin{align}
J^{\tilde{q}\tilde{q},b;\text{S}}(\varepsilon) &= \frac{1}{\varepsilon }\left(1- d_2 \ln(1+C)\right) \\
J^{\tilde{q}\tilde{q},b;\text{NS}} &=\frac{1}{C} \ln(1+C) \left(C + 2\right)+\frac{d_2}{2}
   \left(4 \operatorname{Li}_2(-C)+\ln^2(1+C)\right)   +(1-d_2 \ln(1+C)) \ln\left(\frac{ s}{\bar{s}
   (1-x_0)^2}\right)
\end{align}
with
\begin{equation}
d_2 = \frac{\bar{s}}{\sqrt{\lambda_{ab}}}.
\end{equation}

\subsection{Final expression}
The auxiliary cross section for an emitter and spectator both from the initial state can be recast into the form
    \begin{equation}
    \int_{m+1} \dd\sigma^{\text{A}}_{ab} = \sum_{a'} \int_{x_0}^1 \dd{x}  \int_m \left[\dd\sigma^\text{B}_{ab}\left(\tilde{p}_{a}(x)\right) \frac{\mathcal{F}_{\tilde{a}}}{\mathcal{F}_a} \otimes  \mathbf{I}_{m+b,aa'}\left(x;\varepsilon\right)\right] \label{eq:dAIEFS}
\end{equation}
with the insertion operator
\begin{equation}
    \mathbf{I}_{m+b,aa'}\left(x;\varepsilon,\mu^2;\{p_i,m_i\},p_a,p_b\right) = - \frac{\alpha_s}{2\pi} \frac{(4\pi)^\varepsilon}{\Gamma(1-\varepsilon)} \mathbf{T}_b\cdot \mathbf{T}_{a'} \left(\frac{\mu^2}{\bar{s}}\right)^\varepsilon \frac{1}{\mathbf{T}_{a'}^2} \mathcal{V}^{a,a',b}\left(x;\varepsilon\right). \label{eq:IoperatorIEIS}
\end{equation}
There is no sum over all possible spectators $b$ in \cref{eq:IoperatorIEIS} as we only consider two particles in the initial state.
The flavor functions $\mathcal{V}^{a, a',b}$ are given by the integrated dipoles defined in \cref{eq:IIEIS} via
\begin{equation}
   \mathcal{V}^{a, a',b}(x,m_a,m_b;\varepsilon) =  I^{aa',b}(x;\varepsilon).
\end{equation}

\section{Final-state emitter and final-state spectator}
\label{sec:FEFS}
Since the mass of the initial particles does not influence the splitting behavior, the case of final-state emitter and spectator is already fully covered for the massless and massive case in Refs. \cite{Catani:1996vz,Catani:2002}.

\section{Examples and comparison with the phase space slicing method}
\label{sec:comparision}
In this section, we compare the results of NLO SUSY-QCD corrections for the processes $\tilde{\chi}_1^0 \tilde{t}_1\to t g$ and $\tilde{t}_1 \tilde{t}_1\to t t$ obtained with the phase space slicing method \cite{Harris:2001sx} with the ones obtained with the extension of the dipole subtraction method covered in this paper. Both of these processes are part of the dark matter precision tool \texttt{DM@NLO} which provides NLO and Coulomb corrections for selected (co)-annihilation processes.  
For the first process with a top quark and a gluon in the final state the two-cutoff phase space slicing method is used. Within this approach the three particle phase space is split into a hard and a soft part by imposing a soft cutoff $\delta_s$ on the energy of the radiated gluon. The hard phase space region is split further into a hard and collinear and a hard and non-collinear part through a collinear cutoff $\delta_c$, if the process contains another massless particle:
\begin{equation}
\sigma^{\text{R}}=\sigma^{\text{hard}}_{\text{coll}}(\delta_s,\delta_c)+\sigma^{\text{hard}}_{\text{non-coll}}(\delta_s,\delta_c)+\sigma^{\text{soft}}(\delta_s).
\end{equation}
If there occurs no collinear divergence as in the second process under consideration with two top quarks in the final state, one soft cutoff is sufficient.
In this way the real emission cross section $\sigma^{\text{R}}$ is split into a finite part $\sigma^{\text{hard}}_{\text{non-coll}}(\delta_s,\delta_c)$, which is safe for numerical evaluation in four dimensions, whereas the two other parts have to be integrated analytically in $D=4-2\varepsilon$ dimensions to isolate the infrared poles in $\varepsilon$. 
For the numerical comparison we use the following set of Standard Model parameters \cite{Zyla:2020zbs}
\begin{align}
\begin{aligned}
            m_t&=\SI{173.2}{\giga\electronvolt} 
     &  m_b(m_b)&=\SI{4.18}{\giga\electronvolt} & & \\
    \alpha_{\text{em}}(m_Z)&=\SI{0.00781806}{} & 
    \alpha_{\text{s}}(m_Z)&=\SI{0.1184}{}  &   & \\
    m_Z&=\SI{91.1876}{\giga\electronvolt}
    &  \sin\theta_W&=\SI{0.481}{} 
\end{aligned}
\label{eq:SM_params}
\end{align}
along with the example scenario in the phenomenological MSSM with 19 free parameters (pMSSM-19) displayed in \cref{tab:scenario}, where all input parameters are defined at the scale $Q_{\rm SUSY}$, which is also taken to be the renormalization scale $\mu_R = Q_{\rm SUSY}$.
The associated physical mass spectrum is computed with the public spectrum generator \texttt{SPheno 3.3.3} \cite{Porod:2003um,Porod:2011nf}. The most relevant masses for the two given processes such as the mass of the lightest neutralino, the lightest stop and the gluino are shown in \cref{tab:scenario} as well. 
\begin{table}[h]
    \begin{center}
    \begin{tabular}{cccccccccccc}
    \toprule
          $M_1$ & $M_2$ & $M_3$ & $M_{\tilde{l}_L}$ & $M_{\tilde{\tau}_L}$ & $M_{\tilde{l}_R}$ & $M_{\tilde{\tau}_R}$ & $M_{\tilde{q}_L}$ & $M_{\tilde{q}_{3L}}$ & $M_{\tilde{u}_R}$ \\  
    \midrule      
          1278.5 & 2093.5 & 1267.2 & 3134.1 & 1503.9 & 2102.5 & 1780.4 & 3796.6 & 2535.1 & 3995.0 \\
    \midrule
          $M_{\tilde{t}_R}$ & $M_{\tilde{d}_R}$ & $M_{\tilde{b}_R}$ & $A_t$ & $A_b$ & $A_\tau$ & $\mu$ & $m_{A^0}$ & $\tan\beta$ & $Q_{\rm SUSY}$ \\
    \midrule
     1258.7 & 3133.2 & 3303.8 & 2755.3 & 2320.9 & -1440.3  & -3952.6 & 3624.8 & 15.5 & 1784.6\\
    \bottomrule 
    \end{tabular}
    \\[2mm]
    \begin{tabular}{cccccccc}
    \toprule
    $m_{\tilde{\chi}^0_1}$ & $m_{\tilde{\chi}^0_2}$ & $m_{\tilde{\chi}^{\pm}_1}$ & $m_{\tilde{t}_1}$ & $m_{\tilde{b}_1}$ & $m_{\tilde{g}}$ & $m_{h^0}$ & $m_{H^0}$ \\
    \midrule
    1279.7 & 2153.6 & 2153.5 & 1301.9 & 2554.2 & 1495.5 & 125.8 & 3625.6 \\ 
    \bottomrule
    \end{tabular}
    \end{center}
    \caption{Reference scenario within the pMSSM-19 and the corresponding physical mass spectrum for the numerical comparison. All dimensionful quantities are given in \si{\giga\electronvolt}. }
    \label{tab:scenario}
\end{table}
We emphasize that the parameters in \cref{eq:SM_params} and \cref{tab:scenario} undergo changes through the renormalization scheme defined in Refs. \cite{Harz:2012fz,Harz:2014tma,Schmiemann:2019czm}.
For all considered processes the integration of the three particle phase space and of the "plus"-distribution within the dipole subtraction method is performed with the \emph{Vegas} algorithm from the \texttt{CUBA} library \cite{Hahn:2004fe}, whereas the two particle phase space is integrated with a non-adaptive Gauss-Kronrod-Patterson integrator adapted from \texttt{FormCalc} \cite{Hahn:2004rf}. Both algorithms also provide an estimate on the numerical error. These are combined to the total numerical error of the NLO correction 
\begin{equation}
    \varepsilon_{\text{NLO}} = \sqrt{\varepsilon^2_{\text{plus}} + \varepsilon^2_{\text{V}} + \varepsilon^2_{\text{R}}},
    \label{eq:error}
\end{equation}
which is computed as the geometric mean of the respective numerical errors of the "plus"-distribution $\left(\varepsilon_{\text{plus}}\right)$, the virtual $\left(\varepsilon_{\text{V}}\right)$ and the real $\left(\varepsilon_{\text{R}}\right)$ contribution. For the PSS approach, $\varepsilon_{\text{plus}}$ is set to zero. 

\subsection{The process $\tilde{\chi}_1^0 \tilde{t}_1\to t g$}
The $\order{\alpha_s}$ SUSY-QCD corrections to neutralino-stop coannihilation into a gluon and a top quark have been discussed in Ref. \cite{Harz:2014tma} including a detailed account on the application of the phase space slicing method with two cutoffs. \footnote{Note that the numerical results in the present paper cannot be directly compared to those in Fig. 10 (lower right) of Ref. \cite{Harz:2014tma}, as a term $\nicefrac{-\pi^2}{3}$ from the expansion of $\nicefrac{\Gamma (1-\varepsilon )}{\Gamma (1-2 \varepsilon ) \Gamma (1+\varepsilon )}$ in the (correct) Eq. (2.28) was missing in the numerical implementation. Replacing $A_0^{g\to gg } \to A_0^{g\to gg } -  \frac{r}{6} C_A$ with $r=1$ in Eq. (2.37) as required for dimensional reduction is, however, numerically insignificant.}

The process $\tilde{\chi}^0_1 \tilde{t}_1\to t g$ receives contributions at next-to-leading order from the two real emission processes 
\begin{equation}
    \tilde{t}_1(p_a) + \tilde{\chi}_1^0(p_b) \longrightarrow t(p_1) +g(p_2) +g(p_3)
\end{equation}
and
\begin{equation}
     \tilde{t}_1(p_a) + \tilde{\chi}_1^0(p_b)\longrightarrow t(p_1) +q(p_2)+ \bar{q}(p_3).
\end{equation}
The decay of a gluon into a massless quark-antiquark pair has to be included, since the first four quark flavors $N_f=4$ are treated as effectively massless in \texttt{DM@NLO}.
For a process involving only three colored particles, the different color projections fully factorize in terms of the associated quadratic Casimirs. Therefore, it is not necessary to calculate any color-correlated tree amplitudes thanks to the relation
\begin{equation}
2 \mathbf{T}_2 \cdot \mathbf{T}_3  \ket{1,2,3} =\left(\mathbf{T}_1^2 -\mathbf{T}_2^2 -\mathbf{T}_3^2 \right) \ket{1,2,3},
\end{equation}
which holds analogously for $\mathbf{T}_1 \cdot \mathbf{T}_3$ and $\mathbf{T}_1 \cdot \mathbf{T}_2$. The dipole factorization formula in \cref{eq:dipole_formula} yields a total of ten dipoles to compensate all infrared divergences in the three-particle phase space for the process with two final-state gluons
\begin{align} 
\mathcal{D}_{31,2}&=\frac{1}{2 p_1\cdot p_3} \frac{C_A}{2 C_F}\langle \mathbf{V}_{g_3 t_1,2}\rangle |\mathcal{M}_2\left(p_a,\tilde{p}_{31},\tilde{p}_2\right)|^2  \\
\mathcal{D}_{21,3}&=\frac{1}{2 p_1\cdot p_2} \frac{C_A}{2 C_F} \langle \mathbf{V}_{g_2 t_1,3}\rangle |\mathcal{M}_2\left(p_a,\tilde{p}_{21},\tilde{p}_3\right)|^2 \\
\mathcal{D}_{23,1}&=\frac{1}{2 p_2\cdot p_3} \frac{1}{2}\langle\mu| \mathbf{V}_{g_2 g_3,1}|\nu\rangle \mathcal{T}_{\mu\nu}\left(p_a,\tilde{p}_{1},\tilde{p}_{23}\right) \label{eq:gluon_dipole} \\
\mathcal{D}_{23}^a&=\frac{1}{2 p_2\cdot p_3} \frac{1}{x_{23,a}} \frac{1}{2}\langle\mu| \mathbf{V}_{g_2 g_3}^a |\nu \rangle  \mathcal{T}_{\mu\nu}\left(\tilde{p}_a,p_1,\tilde{p}_{23}\right) \\
\mathcal{D}_{31}^a&=\frac{1}{2 p_1\cdot p_3} \frac{1}{x_{31,a}} \left(1- \frac{C_A}{2 C_F}\right) \langle \mathbf{V}_{g_3 t_1}^a\rangle |\mathcal{M}_2\left(\tilde{p}_a,\tilde{p}_{31},p_2\right)|^2   \\
\mathcal{D}_{21}^a&=\frac{1}{2 p_1\cdot p_2}  \frac{1}{x_{21,a}}   \left(1- \frac{C_A}{2 C_F}\right) \langle \mathbf{V}_{g_2 t_1}^a\rangle |\mathcal{M}_2\left(\tilde{p}_a,\tilde{p}_{21},p_3\right)|^2 \\
\mathcal{D}^{a3}_2&= \frac{1}{2 p_a\cdot p_3}  \frac{1}{x_{32,a}}  \frac{C_A}{2 C_F}\langle \mathbf{V}^{\tilde{t}_{1,a} g_3}_2 \rangle |\mathcal{M}_2\left(\tilde{p}_{a3},p_1,\tilde{p}_2\right)|^2   \\\
\mathcal{D}^{a2}_3&= \frac{1}{2 p_a\cdot p_2} \frac{1}{x_{23,a}} \frac{C_A}{2 C_F} \langle \mathbf{V}^{\tilde{t}_{1,a} g_2}_3 \rangle |\mathcal{M}_2\left(\tilde{p}_{a2},p_1,\tilde{p}_3\right)|^2 \\
\mathcal{D}^{a3}_1&=\frac{1}{2 p_a\cdot p_3} \frac{1}{x_{31,a}}  \left(1- \frac{C_A}{2 C_F}\right)  \langle \mathbf{V}^{\tilde{t}_{1,a} g_3}_1 \rangle |\mathcal{M}_2\left(\tilde{p}_{a3},\tilde{p}_1,p_2\right)|^2     \\
\mathcal{D}^{a2}_1&=\frac{1}{2 p_a\cdot p_2} \frac{1}{x_{21,a}}  \left(1- \frac{C_A}{2 C_F}\right)  \langle \mathbf{V}^{\tilde{t}_{1,a} g_2}_1 \rangle |\mathcal{M}_2\left(\tilde{p}_{a2},\tilde{p}_1,p_3\right)|^2  
\end{align}
with the tree level matrix element squared $|\mathcal{M}_2\left(p_{\tilde{t}_1},p_t,p_g\right)|^2$.
The tensor $\mathcal{T}_{\mu\nu}$ corresponds to the leading order squared amplitude where the polarization vector $\epsilon^\mu_{\lambda}\left(\tilde{p}_{ij}\right)$ of the emitter gluon has been amputated. Since both gluons can become soft in the splittings $\tilde{t}_1\to \tilde{t}_1 g$ and $t\to t g$, one dipole is introduced for each individual gluon in the final state. 
To cancel the collinear divergences from the production of the $N_f$ massless quark-antiquark pairs, the dipoles  
\begin{align}
\mathcal{D}_{23,1} &=\frac{1}{2 p_2\cdot p_3} \frac{1}{2}   \langle\mu|  \mathbf{V}_{q_2 \overline{q}_3,1}|\nu\rangle \mathcal{T}_{\mu\nu}\left(p_a,\tilde{p}_{23},\tilde{p}_1\right) \label{eq:dipole_QQbar_1}\\ 
\mathcal{D}_{23}^a &=\frac{1}{2 p_2\cdot p_3} \frac{1}{x_{23,a}} \frac{1}{2} \langle\mu| \mathbf{V}_{q_2 \overline{q}_3}^a|\nu\rangle \mathcal{T}_{\mu\nu}\left(\tilde{p}_a,p_1,\tilde{p}_{23}\right) \label{eq:dipole_QQbar_2}
\end{align}
are needed. 
The auxiliary cross section that cancels the infrared divergences of the virtual one-loop corrections is constructed from the three insertion operators
\begin{multline}
\bra{1,2,3}\mathbf{I}_2(\varepsilon,\mu^2,\{p_i,m_i\})\ket{1,2,3} = \frac{\alpha_s}{4\pi} \frac{(4 \pi)^\varepsilon}{\Gamma(1-\varepsilon)}  |\mathcal{M}_2|^2 \\ \times \left[ C_A  \left(\frac{\mu^2}{s_{12}}\right)^\varepsilon \left(2 \mathcal{V}^{\text{(S)}}\left(s_{12},m_t,0;\varepsilon\right) +\mathcal{V}^{\text{(NS)}}_g\left(s_{12},0,m_t;\kappa\right) + \mathcal{V}^{\text{(NS)}}_t\left(s_{12},m_t,0\right)-\frac{2\pi^2}{3} \right) \right. \\ \left.
 +  \Gamma^{\textsc{FDH}}_g(\varepsilon)  + \gamma_g\ln\left(\frac{\mu^2}{s_{12}}\right)  + \gamma_g + K_g  + \frac{C_A}{C_F} \left(\Gamma_t\left(\mu,m_t;\varepsilon\right) + \gamma_t \ln(\frac{\mu^2}{s_{12}}) +\gamma_t +K_t\right)\right], \label{eq:exIFEFS}
\end{multline}
\begin{multline}
\bra{1,2,3}\mathbf{I}_{2,\tilde{t}_1}  \left(x;\varepsilon,\mu^2;\{p_i,m_i\},p_a\right)\ket{1,2,3} = \frac{\alpha_s}{4\pi} \frac{(4 \pi)^\varepsilon}{\Gamma(1-\varepsilon)}  |\mathcal{M}_2|^2 \\ \times \left(\left(\frac{\mu^2}{-\bar{\tilde{t}}}\right)^\varepsilon \mathcal{V}_g(x,\tilde{t},0;\varepsilon)+\left(2 - \frac{C_A}{ C_F}\right)\left(\frac{\mu^2}{-\bar{\tilde{u}}}\right)^\varepsilon \mathcal{V}_t(x,\tilde{u},m_t;\varepsilon)\right), \label{eq:exIFEIS}
\end{multline}
\begin{multline}
    \bra{1,2,3}\mathbf{I}_{2,\tilde{q}\tilde{q}}\left(x;\varepsilon,\mu^2;\{p_i,m_i\},p_a\right)\ket{1,2,3} = \frac{\alpha_s}{4\pi} \frac{(4 \pi)^\varepsilon}{\Gamma(1-\varepsilon)}  |\mathcal{M}_2|^2 \\ \times \left(\frac{C_A}{C_F}\left(\frac{\mu^2}{-\bar{\tilde{t}}}\right)^\varepsilon \mathcal{V}^{\tilde{q}, \tilde{q}}(x,\tilde{t},0;\varepsilon)+\left(2 - \frac{C_A}{ C_F}\right)\left(\frac{\mu^2}{-\bar{\tilde{u}}}\right)^\varepsilon \mathcal{V}^{\tilde{q}, \tilde{q}}(x,\tilde{u},m_t;\varepsilon)\right) \label{eq:exIIEFS}
\end{multline}
with $s_{12}= s - m_t^2$ where the first one in \cref{eq:exIFEFS} corresponds to emitter and spectator both from the final state, the second one in \cref{eq:exIFEIS} to final-state emitter with a spectator from the initial state and the last one in \cref{eq:exIIEFS} to an initial-state emitter with final-state spectators. The dipole invariants $\tilde{t}=(\tilde{p}_{\tilde{t}_1} - \tilde{p}_g)^2$ and $\tilde{u}=(\tilde{p}_{\tilde{t}_1} - \tilde{p}_t)^2$ correspond to the Mandelstam variables $t=(p_{\tilde{t}_1}-p_g)^2$ and $u=(p_t-p_{\tilde{t}_1})^2$ in the squared Born amplitude and they play the role of $Q^2$. With that, the "barred variables" $\bar{\tilde{t}}$ and $\bar{\tilde{u}}$ are given by 
\begin{align}
    \bar{\tilde{t}} = \tilde{t} - m^2_{\tilde{t}_1}, \ \ \ \
    \bar{\tilde{u}} = \tilde{u} - m^2_{\tilde{t}_1} - m^2_{t}.
\end{align}
The insertion operator in \cref{eq:exIFEFS} for emitter and spectator both from the final state as well as the related flavor functions
\begin{equation}
\mathcal{V}^{(\text{S})}\left(s_{12},m_t,0;\varepsilon\right) = \frac{1}{2\varepsilon^2} +  \frac{1}{2\varepsilon} \ln\left(\frac{m_t^2}{s_{12}}\right)
- \frac{1}{4} \ln^2\left(\frac{m_t^2}{s_{12}}\right) - \frac{\pi^2}{12}    - \frac{1}{2} \ln\left(\frac{s_{12}}{s}\right) \left[\ln\left(\frac{m_t^2}{s_{12}}\right)+ \ln\left(\frac{m_t^2}{s}\right)\right]
\end{equation}
\begin{multline}
\mathcal{V}^{(\text{NS})}_g\left(s_{12},0,m_t;\kappa\right) =\frac{\gamma_g}{C_A}\left( \ln\left(\frac{s_{12}}{s}\right)-2 \ln\left(\frac{\sqrt{s}-m_t}{\sqrt{s}}\right)- \frac{2 m_t}{\sqrt{s}+m_t}  \right) + \frac{\pi^2}{6}  \\
-  \operatorname{Li}_2\left(\frac{s_{12}}{s}\right) +\left(\kappa-\frac{2}{3}\right) \frac{m_t^2}{s_{12}} \left(\left(2 N_f \frac{T_F}{C_A} - 1\right) \ln\left(\frac{2 m_t}{\sqrt{s}+m_t}\right)\right)  \label{eq:VNSgFEFS}
\end{multline}
\begin{equation}
\mathcal{V}^{(\text{NS})}_t\left(s_{12},m_t,0\right)= \frac{3}{2} \ln\left(\frac{s_{12}}{s}\right) + \frac{\pi^2}{6} - \operatorname{Li}_2\left(\frac{s_{12}}{s}\right) - 2 \ln\left(\frac{s_{12}}{s}\right) - \frac{m_t^2}{s_{12}} \ln\left(\frac{m_t^2}{s}\right)
\end{equation}
are provided in Ref. \cite{Catani:2002} where the function $\Gamma_j$ for massive quarks reads
\begin{align}
\Gamma_t\left(\mu,m_t;\varepsilon\right)&=C_F\left(\frac{1}{\varepsilon} + \frac{1}{2} \ln\left(\frac{m_t^2}{\mu^2}\right) - 2\right).
\end{align}
The value of the variable $\kappa$ in \cref{eq:VNSgFEFS} can be chosen arbitrarily as its dependence must cancel out between the virtual and real part. Within the numerical comparison it is set to $\kappa=0$.
Note that due to Bose symmetry the dipoles which are related through the interchange of an emitted gluon result in the same integrated dipole. Therefore, it is sufficient to incorporate one of the integrated counterparts and weight it with a factor of two which gets cancelled by the Bose symmetry factor $S_3=\frac{1}{2}$ of the associated real emission cross section. This counting of symmetry factors is already incorporated into the definition of the flavor functions $\mathcal{V}_j$\footnote{The counting of symmetry factors for the general case of going from $m+1$ to $m$ particles for a gluon and quark as emitter is discussed extensively in Ref. \cite{Catani:1996vz}.}.
In order to perform the convolution in \cref{eq:PSfactorizFEIS}, the well-known parametrization of the two-particle phase 
\begin{equation}
    \int \dd\phi\left(P(x),p_k;p_a+p_b\right) = \frac{1}{(4\pi)^2}
\frac{1}{\sqrt{\lambda\left(s,m_a^2,m_b^2\right)}} \int_{Q^2_{-}\left(x\right)}^{Q^2_{+}\left(x\right)} \dd{Q^2} \int_0^{2\pi} \dd\varphi_k \label{eq:int_cosTheta}
\end{equation}
is inserted, where $\varphi_k$ denotes the azimuthal angle of $p_k$ in the center-of-mass system of $p_a+p_b$. Since the integrand is rotationally invariant, the integration over $\varphi_k$ yields a factor of $2\pi$. We still need to determine the  integration limits of $Q^2$ as a function of $x$ which is achieved by expressing $Q^2$ in the c.m. frame of $p_b$ and $p_k$
\begin{align}
Q^2 &= m_b^2+m_k^2  -2 E_b E_k + 2|\vec{p}_b||\vec{p}_k|\cos\vartheta \nonumber \\
&=m_b^2 + m_k^2 - \frac{\left(s + m_b^2 - m_a^2\right) \left(s + m_k^2 - P^2\right)}{2 s} +\cos\vartheta 
  \frac{\sqrt{\lambda\left(s,m_a^2,m_b^2\right)}  \sqrt{\lambda\left(s,m_k^2,P^2\right)}}{2 s}  \label{eq:Q2expanded}
\end{align}
where $\vartheta$ corresponds to the angle between $\vec{p}_b$ and $\vec{p}_k$. The $x$ dependence enters by expressing $P^2$ through $x$ and $Q^2$ as given in \cref{eq:relation_P2Q2x} which results in an equation which we can solve for $Q^2$. The integration limits
\begin{equation}
    Q^2_{\pm}(x) = \frac{1}{2}\frac{\alpha(x)\pm\beta(x)}{x s + (1-x) (m_b^2 -x m_a^2)}
\end{equation}
with the abbreviations 
\begin{multline}
    \alpha(x) = x^2
   \left(m_a^4+2 m_a^2 (m_b^2+m_k^2)-(m_b^2-s)^2\right)+2 m_b^2 (m_a^2+m_j^2) \\ -x \left(m_a^4+m_a^2 (4 m_b^2+m_j^2+m_k^2-s)-(m_b^2-s)
    (m_b^2-m_j^2-m_k^2)\right),
\end{multline}
\begin{multline} 
    \beta(x)=x \sqrt{\lambda\left(m_a^2,m_b^2,s\right)} \sqrt{\left(m_a^2-m_b^2\right)^2 (1-x)^2+(1-x)\left(2 m_a^2 \left(m_j^2+m_k^2 (2 x-1)-s x\right)\right.} \\  \overline{\left.-2 m_b^2 \left(m_j^2+m_k^2-s x\right)\right)+\lambda(x s,m_k^2,m_j^2)}
\end{multline}
are then obtained by setting $\cos\vartheta$ to its extreme values $-1$ and $1$. Within the integration over $Q^2$ two different kinematical configurations have to be distinguished. The variable $\tilde{t}$ in \cref{eq:exIFEIS,eq:exIIEFS} equals $Q^2$ for the cases $m_j=0$, $m_k=m_t$ whereas $\tilde{u}$ equals $Q^2$ for $m_j=m_t$, $m_k=0$. After having fixed the values of $x$ and $Q^2$ ($\tilde{u}$ and $\tilde{t}$) in the phase space integration, the squared c.m. energy $\tilde{s}$ of the new initial state with momenta $\tilde{p}_a$ and $p_b$ can be determined as
\begin{multline} 
    \tilde{s}= (\tilde{p}_a + p_b)^2 = m_a^2 + m_b^2 + \frac{1}{ R(x)} \left[x(s-m_a^2-m_b^2) \right. \\ \left. + \frac{\bar{Q}^2+2 m_a^2 x}{2Q^2} \left(m^2_b-m_k^2+Q^2\right)\right] -\frac{Q^2 + m_a^2 -m_j^2}{2 Q^2} \left(m_b^2 - m_k^2+Q^2\right).
\end{multline}
The remaining "dipole Mandelstam variable" $\tilde{u}$ for $Q^2=\tilde{t}$ and vice versa can then be deduced from $\tilde{s}+\tilde{u}+\tilde{t}=m_a^2+m_b^2+m_j^2+m_k^2$. As the squared tree-level matrix element is a function of the usual Mandelstam variables $s$, $t$ and $u$, we only need to substitute those through the dipole invariants $\tilde{s}$, $\tilde{t}$ and $\tilde{u}$, respectively, in order to formulate the tree-level matrix element in terms of the dipole momenta.

\begin{figure}[h]
    \centering
	\begin{subfigure}[c]{0.485\textwidth}		
		\centering	
		\includegraphics[width=\textwidth]{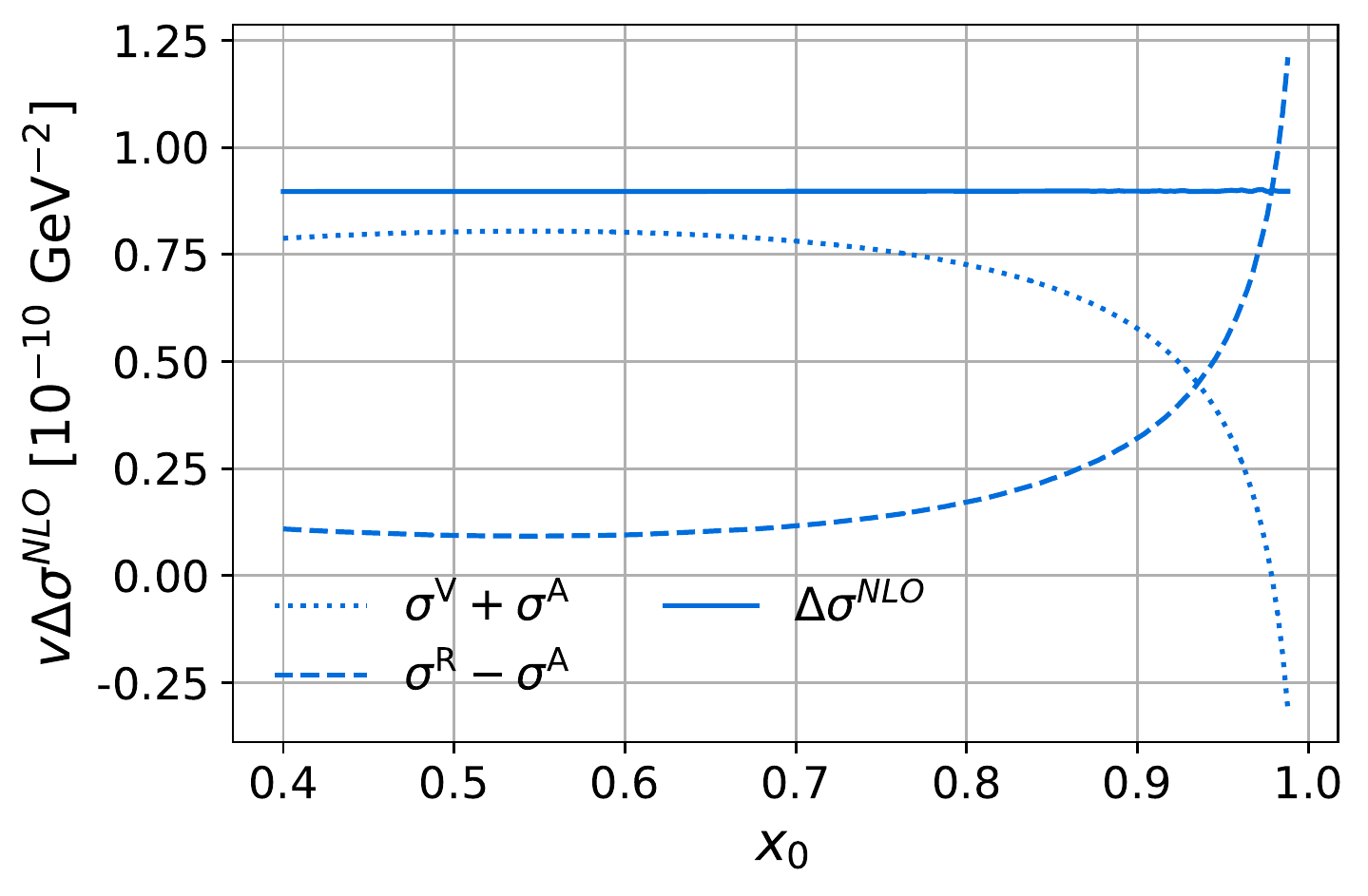}
	\end{subfigure}
	\begin{subfigure}[c]{0.495\textwidth}
		\centering
		\includegraphics[width=\textwidth]{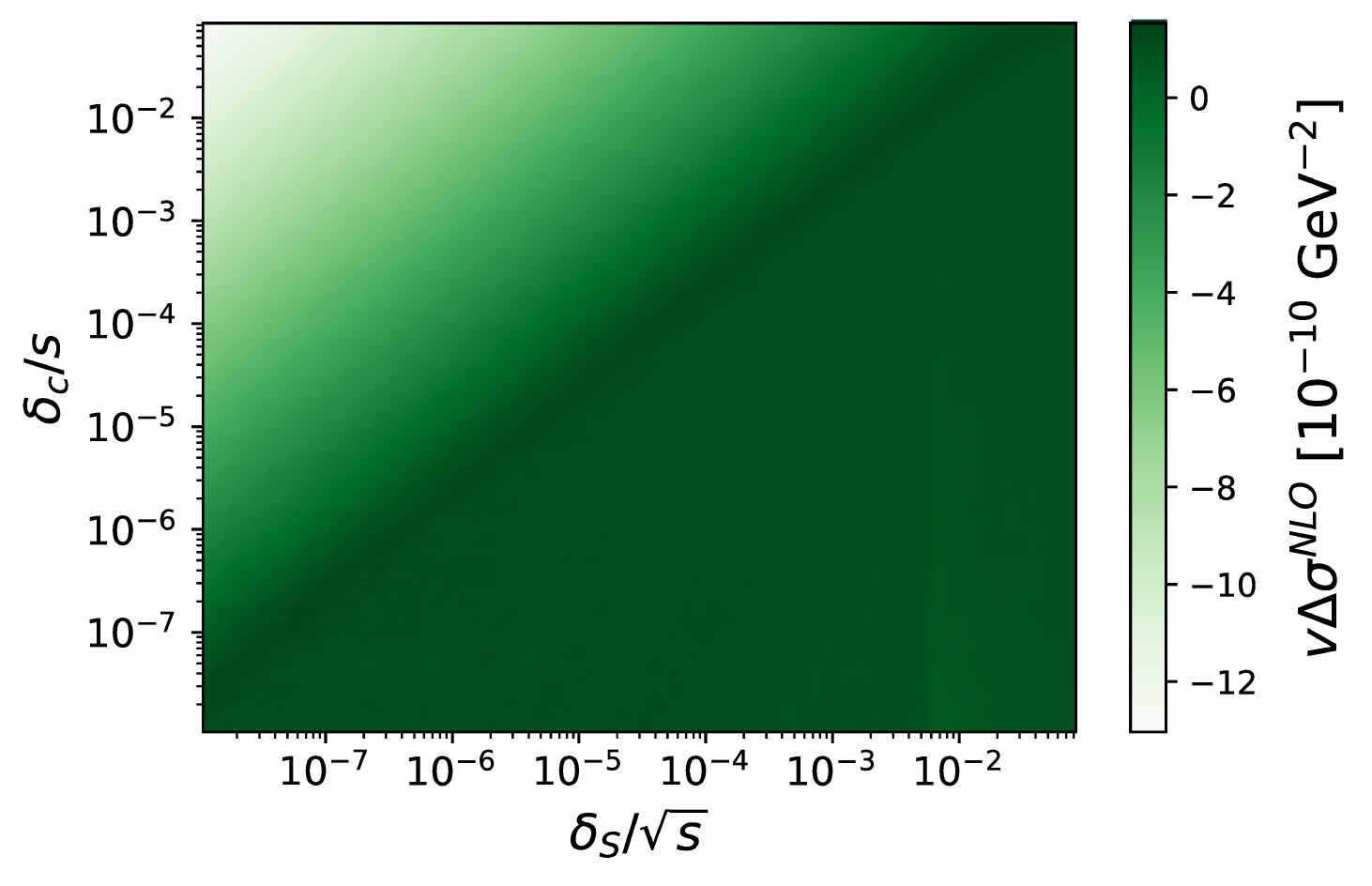}
	\end{subfigure}
    \caption{The NLO correction times velocity $v\Delta\sigma^{\text{NLO}}$ subdivided into the virtual part plus the auxiliary cross section $\sigma^{\text{V}}+\sigma^{\text{A}}$ and the real part minus the auxiliary cross section $\sigma^{\text{R}}-\sigma^{\text{A}}$ for the process $\tilde{\chi}_1^0 \tilde{t}_1\to t g$ for different values of the lower integration limit $x_0$ (left) as well as the dependence of the NLO correction obtained with the slicing method on the soft $\delta_s$ as well as the collinear cutoff $\delta_c$ (right). Both plots are created for the c.m. momentum $p_{\mathrm{cm}}=\SI{100}{\giga\electronvolt}$. }
     \label{fig:cutoff_gluon}
\end{figure}

The independence of the final result on the lower integration limit $x_0$ is shown in \cref{fig:cutoff_gluon}. 
For the numerical comparison the value $x_0=0.9$ was chosen as it fulfills the condition in \cref{eq:x0Cond} for all probed c.m. momenta.
For the determination of appropriate values for the soft and collinear cutoff, the behavior of the NLO correction is examined in dependence of both, which is shown in \cref{fig:cutoff_gluon}. The cutoffs are chosen to be $p_2^0,p_3^0 \geq\delta_s=3.0\cdot 10^{-4}\sqrt{s}$ and $2 p_2\cdot p_3\geq\delta_c =3.0\cdot 10^{-6} s$ such that they are located in the broad plateau region in the lower right half of the plot. 

\begin{figure}[h]
    \centering
    \includegraphics[width=0.75\textwidth]{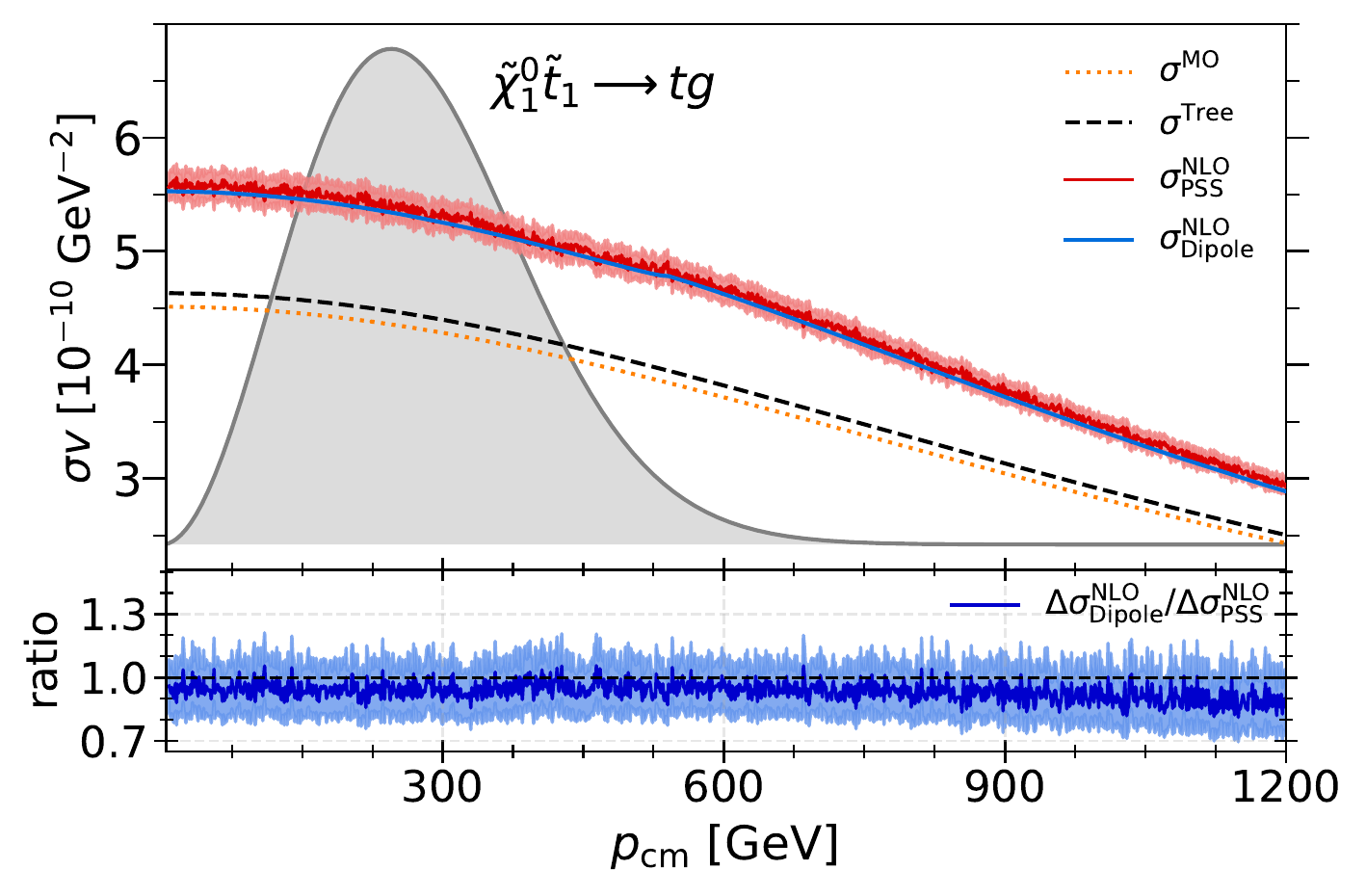}
    \caption{Neutralino-stop coannihilation cross section $\sigma v$ with a top and a gluon in the final state for the example scenario defined in \cref{tab:scenario}. The leading order result is computed with \texttt{MicrOMEAGs 2.4.1} \cite{Belanger:2001fz,Belanger:2006is} (MO) and \texttt{DM@NLO} (Tree). The NLO results are calculated with the phase space slicing method (PSS) and the dipole method (Dipole). The lower panel shows the ratio of of the NLO corrections obtained with the two different approaches. The uncertainty band in the upper panel corresponds to the total numerical error $\varepsilon_{\text{NLO}}$ defined in \cref{eq:error}. The gray shaded area shows the thermal velocity distribution of the neutralino at the freeze-out temperature in arbitrary units.}
    \label{fig:XSec_gluon} 
\end{figure}

In \cref{tab:gluonComp} and \cref{fig:XSec_gluon} the total cross section obtained with the two different methods is given for c.m. momenta $p_{\mathrm{cm}}$, that are typical for dark matter annihilation. Even though all chosen cutoffs for a momentum of $\SI{100}{\giga\electronvolt}$ lie in the plateau region shown in the right plot of \cref{fig:cutoff_gluon}, the central values of the correction for the smallest and largest cutoff differ by $\SI{13}{\percent}$, while the dependence on the artificially introduced lower integration limit $x_0$ of the dipole method is completely compensated between the virtual and real part. Furthermore, the total numerical error of the result obtained with the phase space slicing method for the NLO correction increases with decreasing cutoff values which is expected as the real cross section blows up like $\ln({\delta_s}/{s})$ in the soft region and like $\ln({\delta_c}/{s})$ in the collinear one. In addition, the integration error of the dipole method is at least one order of magnitude lower than the one of the slicing method so that the error of the dipole result is smaller than the linewidth in the plot. Both of these findings, the cutoff dependence as well as the integration error, show the superiority of the dipole subtraction method with respect to precision. 
\begin{table}[h]
    \begin{center}
    \begin{tabular}{cccccc}
    \toprule
      $p_{\mathrm{cm}}$ [\si{\giga\electronvolt}] & $v\sigma^{\text{Tree}}$  & Method & $\delta_s/\sqrt{s}$ & $\delta_c/s$ & $v\Delta\sigma^{\text{NLO}}$ \\
    \midrule
     \multirow{4}{*}{100} & \multirow{4}{*}{4.604596} & & $10^{-2}$ & $10^{-3}$ & $0.915 \pm 0.036$  \\
      & &  PSS & $10^{-4}$ & $10^{-6}$ & $0.974 \pm 0.152$  \\
     &  & & $10^{-6}$ & $10^{-7}$ & $1.033 \pm 0.241$ \\
     \cmidrule{3-6}
     &  & Dipole &  &  & $0.891\pm 0.002$ \\
     \midrule
      \multirow{4}{*}{1200} & \multirow{4}{*}{2.501535} & & $10^{-2}$ & $10^{-3}$ & $0.408 \pm 0.021$ \\
      & &  PSS & $10^{-4}$ & $10^{-6}$ & $0.429 \pm 0.083$ \\
     &  & & $10^{-6}$ & $10^{-7}$ & $0.458 \pm 0.135$ \\
     \cmidrule{3-6}
     &  & Dipole &  &  & $0.385 \pm 0.001$ \\
    \bottomrule 
    \end{tabular}
    \end{center}
    \caption{Results on the correction $v\Delta\sigma^{\text{NLO}}$ of the process $\tilde{\chi}_1^0 \tilde{t}_1\to t g$ for two different c.m. momenta $p_{\mathrm{cm}}$. All cross sections times velocity are given in  $10^{-10}$ \si{\giga\electronvolt^{-2}}.}
    \label{tab:gluonComp}
\end{table}

\subsection{The process $\tilde{t}_1 \tilde{t}_1\to t t$}
As another example, the process
\begin{equation}
    \tilde{t}_1(p_a,s)+\tilde{t}_1(p_b,t) \longrightarrow t(p_1,i)+t(p_2,j)+g(p_3,a)
\end{equation}
is considered where the parentheses contain the particle momenta $p_a$, $p_b$, $p_1$, $p_2$, $p_3$ and the corresponding color indices $s$, $t$, $i$, $j$, $a$. This process is chosen as it allows to demonstrate and compare the dipole formalism for situations with two massive and color charged particles in the initial state. The next-to-leading order corrections for this process performed with the slicing method are discussed in Ref. \cite{Schmiemann:2019czm}. The auxiliary squared matrix element receives contributions from in total twelve dipoles and reads
\begin{equation}
    \left|\mathcal{M}^{\text{A}}_{3}\right|^2 = \mathcal{D}_{13,2} + \mathcal{D}_{23,1}+ \mathcal{D}_{13}^a +\mathcal{D}_{13}^b+\mathcal{D}_{23}^a+\mathcal{D}_{23}^b+ \mathcal{D}^{a3}_1 + \mathcal{D}^{b3}_1 + \mathcal{D}^{a3}_2+ \mathcal{D}^{b3}_2 +\mathcal{D}^{a3,b} +\mathcal{D}^{b3,a},
\end{equation}
where the subtraction functions are consistently set to zero for values of $x$ below $x_0 = \frac{2 m_t^2}{\bar{s}}$ in conjunction with \cref{eq:xhatIEIS}.

\begin{figure}[h]
    \centering
    \includegraphics[width=0.75\textwidth]{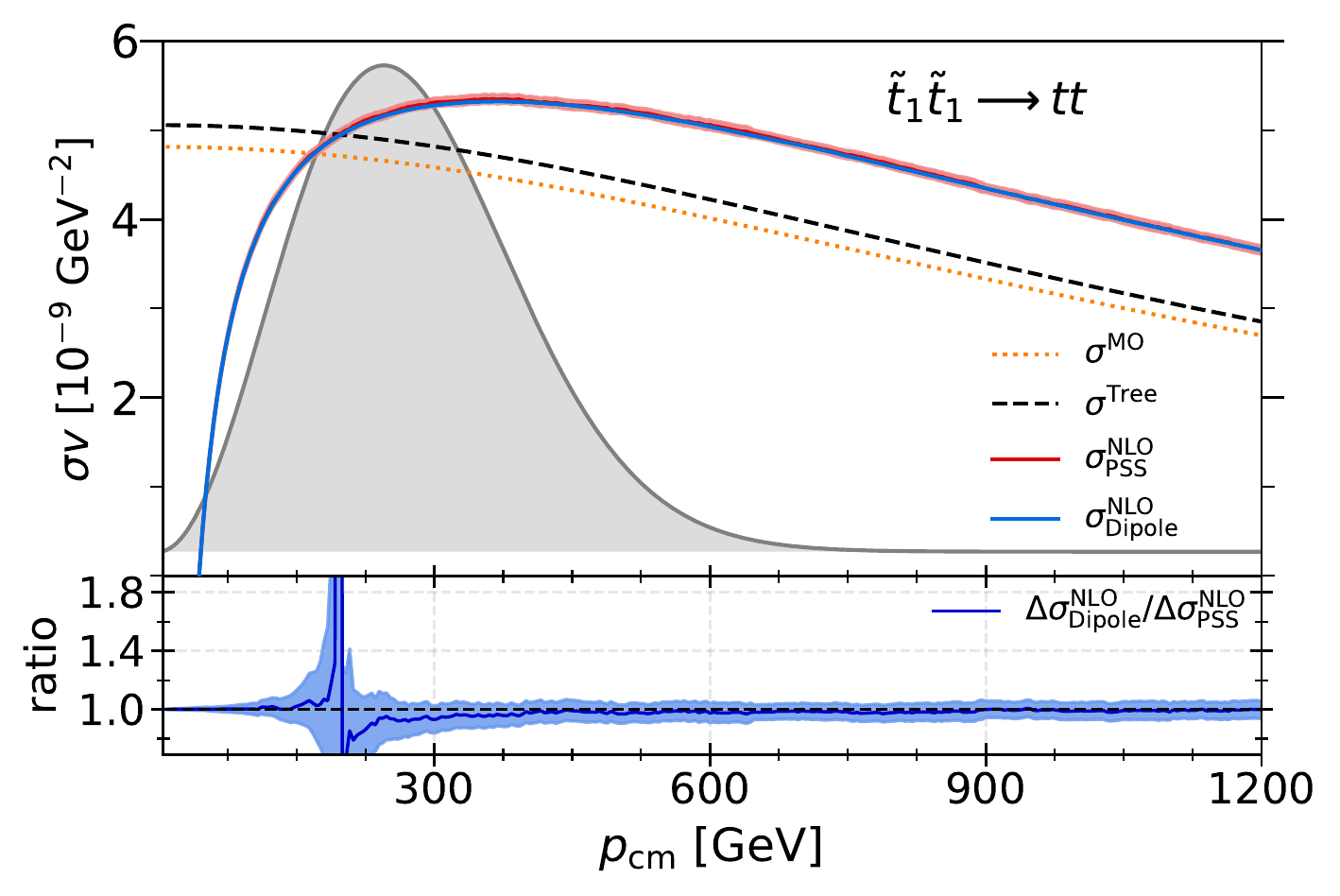}
    \caption{Same as \cref{fig:XSec_gluon} for the annihilation process $\tilde{t}_1 \tilde{t}_1\to t t$.}
    \label{fig:XSec_tt}
\end{figure}

For a process involving four colored particles it is no longer possible to factorize the color charge algebra. However, it follows from color conservation, that four of the six color charge operators $\mathbf{T}_i \mathbf{T}_j$ with $i \neq j$ can be expressed through the quadratic Casimir invariants and $\mathbf{T}_1 \mathbf{T}_2$, $\mathbf{T}_1 \mathbf{T}_3$ giving \cite{Catani:1996vz} 
\begin{align}
\mathbf{T}_3 \mathbf{T}_4 \ket{1,2,3,4} &= \left[\frac{1}{2}\left(C_1+C_2 - C_3-C_4\right) + \mathbf{T}_1 \mathbf{T}_2\right] \ket{1,2,3,4} \\
\mathbf{T}_2 \mathbf{T}_4 \ket{1,2,3,4} &= \left[\frac{1}{2}\left(C_1+C_3 - C_2-C_4\right) + \mathbf{T}_1 \mathbf{T}_3\right] \ket{1,2,3,4}\\
\mathbf{T}_2 \mathbf{T}_3 \ket{1,2,3,4} &= \left[\frac{1}{2}\left(C_4-C_1-C_2-C_3\right) -\mathbf{T}_1 \mathbf{T}_2 - \mathbf{T}_1 \mathbf{T}_3\right] \ket{1,2,3,4} \\
\mathbf{T}_1 \mathbf{T}_4 \ket{1,2,3,4} &= -\left(C_1 + \mathbf{T}_1 \mathbf{T}_2+\mathbf{T}_1 \mathbf{T}_3\right) \ket{1,2,3,4}.
\end{align}
The four color charge operators are associated with the particles in our process as follows:
\begin{equation}
 \mathbf{T}_1 = \mathbf{T}_{\tilde{q}_s}, \ \mathbf{T}_2 = \mathbf{T}_{\tilde{q}_t}, \ \mathbf{T}_3 = \mathbf{T}_{t_i}, \ \mathbf{T}_4 = \mathbf{T}_{t_j}.
\end{equation}
For the remaining two operators the color correlations have to be evaluated explicitly:
\begin{align}
&\bra{1,2,3,4} \mathbf{T}_1 \mathbf{T}_2 \ket{1,2,3,4} =  \left[\mathcal{M}^{ij;lt}_2\right]^\ast  T^c_{s l} T^c_{k t}  \mathcal{M}^{ij;sk}_2, \\
&\bra{1,2,3,4} \mathbf{T}_1 \mathbf{T}_3 \ket{1,2,3,4} = \left[\mathcal{M}_2^{ij;kt}\right]^\ast  \left( -T^c_{s k} T^c_{i l}\right)  \mathcal{M}_2^{lj;st} 
\end{align}
with the tree level matrix element $\mathcal{M}^{ij;st}_2$. As the application of the dipole formulas has already been exemplified in the previous section for all emitter-spectator pairs besides the configuration where both are from the initial state, we only cover the two particle phase space integration in the convolution in \cref{eq:PSIEIS}. In order to provide a general expression for the parametrization of the phase space, the masses related to the momenta $p_1$ and $p_2$ are labelled as $m_1$ and $m_2$ and we distinguish the masses $m_a$ and $m_b$ of the initial particles even though they are identical in this case. Since the variable $x$ enters the phase space integration only through the reduced squared c.m. energy $\tilde{s}$ given in \cref{eq:stildeIEIS}
the well-known parametrization 
\begin{equation}
    \int\dd{\phi}\left(\tilde{p}_{k}(x);\tilde{p}_{ai}(x)+p_b\right) = \frac{1}{(4\pi)^2 \sqrt{\lambda_{\tilde{s}}}}  \int_{q^2_-(\tilde{s})}^{{q^2_+(\tilde{s})}} \dd{q^2} \int_0^{2 \pi} \dd{\varphi'} 
\end{equation} 
with the integration limits
\begin{align}
q^2_{\pm}\left(\tilde{s}\right)&=m_a^2 + m_1^2 - \frac{\left(\tilde{s} + m_a^2 - m_b^2\right) \left(\tilde{s} + m_1^2 - m_2^2\right)}{2 \tilde{s}} \pm 
  \frac{\sqrt{\lambda_{\tilde{s}}}}{2 \tilde{s}} \sqrt{\lambda\left(\tilde{s},m_1^2,m_2^2\right)}
\end{align}
can be employed where $q^2=(\tilde{p}_{a3}-\tilde{p}_1)^2$ plays the role of a Mandelstam variable and the abbreviation $\lambda_{\tilde{s}}$ is given by $\lambda_{\tilde{s}} = \lambda(\tilde{s},m_a^2,m_b^2)$. The remaining dipole Mandelstam variable that enters the squared Born amplitude is determined through $m_a^2 + m_b^2 + m_1^2 + m_2^2 - \tilde{s} -q^2$.
For the numerical comparison in \cref{fig:XSec_tt}, the cutoff for the slicing method is chosen as $p_3^0 \geq\delta_s=10^{-5}\sqrt{s}$. In \cref{tab:stst2tt}, results on the NLO corrections for different cutoff values $\delta_s$ and c.m. momenta are shown in comparison with the result of the dipole approach. Similar to the previous example, the integration error of the slicing method increases with decreasing cutoff values while the errors of dipole method are at least one order of magnitude lower than the ones for small cutoff values indicating again that the dipole method is ahead of the slicing approach. 

\begin{table}[h]
    \begin{center}
    \begin{tabular}{ccccc}
    \toprule
      $p_{\mathrm{cm}}$ [\si{\giga\electronvolt}] & $v\sigma^{\text{Tree}}$  & Method & $\delta_s/\sqrt{s}$ & $v\Delta\sigma^{\text{NLO}}$ \\
    \midrule
     \multirow{4}{*}{100} & \multirow{4}{*}{5.030288} & & $10^{-2}$  & $-1.392 \pm 0.018$ \\
      & &  PSS & $10^{-4}$  & $-1.407 \pm 0.032$  \\
     &  & & $10^{-6}$  & $-1.399 \pm 0.053$ \\
     \cmidrule{3-5}  
     &  & Dipole &   & $-1.410\pm 0.007$ \\
     \midrule
      \multirow{4}{*}{1200} & \multirow{4}{*}{2.853008} & & $10^{-2}$  & $0.821 \pm  0.016$ \\
      & &  PSS & $10^{-4}$  & $0.810\pm 0.036$ \\
     &  & & $10^{-6}$  & $0.787 \pm 0.062$ \\
     \cmidrule{3-5}
     &  & Dipole &   & $0.802 \pm 0.007$ \\
    \bottomrule 
    \end{tabular}
    \end{center}
    \caption{Results on the correction $v\Delta\sigma^{\text{NLO}}$ of the process $\tilde{t}_1 \tilde{t}_1\to t t$ for two different $p_{\mathrm{cm}}$. All cross sections times velocity are given in  $10^{-9}$ \si{\giga\electronvolt^{-2}}.}
    \label{tab:stst2tt}
\end{table}

\section{Summary and outlook}
\label{sec:summary}
In this paper, we have presented an extension of the dipole subtraction formalism  introduced in Refs. \cite{Dittmaier:1999mb,Catani:1996vz,Catani:2002} to massive initial-state particles for NLO (SUSY)-QCD calculations which allows the analytic cancellation of infrared singularities between the virtual and real corrections. Our results are in particular relevant for precision computations of the dark matter relic density in SUSY and non-SUSY models and should in the future allow for automated calculations of the pertinent higher-order corrections.

We reviewed the dipole subtraction method and its underlying notation as well as the factorization of real emission amplitudes in the soft and collinear limit. From there we constructed the dipole splitting functions for different regularization schemes. Even though it is possible to adopt the corresponding kinematics from Dittmaier, the factorization of the $(m+1)$-particle phase space into a $m$-particle and a dipole phase space had to be performed again in $D$ dimensions. Since the integration of the dipole functions and the extraction of the singular terms with the help of the "plus"-distribution turned out to be rather cumbersome, the associated steps were presented in a detailed way. 
At the end of each section dedicated to one of the three emitter-spectator pairs covered in this paper, the results were collected in effective final formulae for the universal insertion operator which allows to render the virtual part infrared finite. 

In order to illustrate the use of the dipole method, our results were applied to the dark matter (co)-annihilation processes $\tilde{\chi}_1^0 \tilde{t}_1\to t g$ and $\tilde{t}_1 \tilde{t}_1\to t t$. Therefore, a general parametrization of the physical two-particle phase space with arbitrary masses in dependence of the associated convolution variables was provided. The results of the dipole method were compared with those obtained with the phase space slicing method. A significant reduction of the integration error was found for the dipole approach. Similar findings were reported by Dittmaier who compared both methods for electroweak processes and in Ref. \cite{Eynck:2001en} for the process $\gamma^\ast \to Q\bar{Q}$. An application of this method to the annihilation processes $\tilde{t}_1 \tilde{t}^\ast_1 \to g g$ and $\tilde{t}_1 \tilde{t}^\ast_1 \to q \bar{q}$ is in preparation. 

\begin{acknowledgments}
We would like to thank Karol Kovařík for useful discussions. J.H. acknowledges support by the Deutsche Forschungsgemeinschaft (DFG, German Research Foundation) Emmy Noether grant HA 8555/1-1. M.K. thanks the School of Physics at the University of New South Wales in Sydney, Australia for its hospitality and financial support through the Gordon Godfrey visitors program. The work of M.K. was also funded by the DFG through grant KL 1266/10-1, the work by M.K. and L.P.W. through the DFG Research Training Group 2149 "Strong and Weak Interactions - from Hadrons to Dark Matter". M.Y.S. acknowledges support by the DFG under Germany’s Excellence Strategy EXC 2121 ”Quantum Universe” - 390833306. Figures and Feynman diagrams presented in this paper have been generated using
\texttt{MatPlotLib} \cite{Hunter:2007} and \texttt{TikZ-Feynman} \cite{Ellis:2016jkw}.
\end{acknowledgments}

\appendix

\section{Dimensional regularization and reduction schemes}
\label{sec:dimension}
There exist two main dimensional schemes for the calculation of matrix elements at one-loop order, which are dimensional regularization and dimensional reduction. Both have in common that the number of dimensions of all momenta and space-time coordinates is analytically continued to $D\neq 4$ dimensions, whereas there remains some freedom regarding the dimensionality of "internal" and "external" vector bosons. Internal gauge bosons are defined as those that appear in a one-particle irreducible diagram of the virtual corrections or that become soft or collinear in a phase space integral related to the real corrections. External gauge bosons are then defined as all other gauge bosons. In order to formulate the different treatments of internal and external gauge fields in a mathematically consistent and precise way, three different spaces are introduced: the original four-dimensional space (4S), the quasi-four-dimensional space (Q4S) and the quasi-$D$-dimensional space (QDS) as a subspace of Q4S \cite{Stockinger:2005gx}. Following the definitions in Ref. \cite{Signer:2008va}, each of the two main schemes has two subvariants. These are in the case of dimensional regularization the \emph{conventional dimensional regularization scheme} (\textsc{CDR}), where internal and external gauge bosons are treated as $D$-dimensional, and the \emph{’t Hooft-Veltman scheme} (\textsc{HV}), where external gauge bosons live in 4S instead of QDS. Within the two subvariants of dimensional reduction, internal gauge bosons are elements of Q4S, whereas external gluons are strictly four-dimensional in the \emph{four-dimensional helicity scheme} (\textsc{FDH}) and also $D$-dimensional in the \emph{original dimensional reduction scheme} (\textsc{DRED}). In order to guarantee that the final result for the physical cross section is independent of the chosen regularization prescription, the gauge bosons in the tree level matrix element $\mathcal{M}_m$ have to be treated like external gauge bosons in the loop amplitude, whereas the particles in the dipole factor $\dd\mathbf{V}_{\text{dipole}}$ have to be treated as internal particles \cite{Catani:1996pk}. The scheme-dependent terms can thus be parameterized by the number of helicity states $h^{\textsc{RS}}_g=2(1-\varepsilon+r\varepsilon)$ of internal gluons, where  
we introduced the parameter $r$ defined as 
\begin{equation}
    r = \begin{cases} 0, \ \textsc{CDR, HV} \\ 1, \  \textsc{DRED, FDH} \end{cases}
\end{equation}
to distinguish between the different schemes according to the definitions above.

\section{Derivation of the phase space factorization}
In this section, we derive the expressions for the dipole phase space that are used for the analytic integration of the dipole splitting functions. 
\subsection{Final-state emitter and initial-state spectator}
\label{sec:PSFE-IS}
We start with the phase space element for $m+1$ particles in the final state in $D = 4-2\varepsilon$ dimensions which is given by \cite{Byckling}
\begin{multline}
\dd\phi_{m+1}(p_{i}, p_{j}, p_{k};p_a+p_b) = (2\pi)\delta^{(D)}(p_{a}+p_{b}-p_{i}-p_{j}-\sum_{k}p_{k})\dfrac{\dd^{D}p_{i}}{(2\pi)^{D-1}}\delta_{+}(p^{2}_{i}-m^{2}_{i}) \\ \times 
\dfrac{\dd^{D}p_{j}}{(2\pi)^{D-1}}\delta_{+}(p^{2}_{j}-m^{2}_{j})\prod_{k}\dfrac{\dd^{D}p_{k}}{(2\pi)^{D-1}}\delta_{+}(p^{2}_{k}-m^{2}_{k}),
\end{multline}
where the modified Dirac delta distribution contains the Heaviside step function $\theta(x)$ and is defined as $\delta_{+}(p^2 - m^2) = \delta(p^2 - m^2)\theta(p^{0})$. 
The momentum of the spectator is $p_{a}$, while the emitter $\tilde{ij}$ splits into two particles with the momenta $p_{i}$ and $p_{j}$. The momenta of the remaining final state particles other than $i$ or $j$ are labelled as $p_k$.
The $(m+1)$-particle phase space is factorized exactly into a $m$-particle phase space and a two-particle phase space through a convolution of the form
\begin{equation}
    \dd\phi_{m+1}(p_{i}, p_{j}, p_{k};p_a+p_b) = \frac{\dd m^2_P}{2\pi}\ \ \dd\phi_m( P , p_k;p_{a}+p_b) \dd\phi_{2}(p_{i}, p_{j};P),
    \label{eq:trueps3}
\end{equation}
where $m^2_P=P^2$ acts as the squared invariant mass related to the momentum $P=p_i+p_j$. As the dipole splitting functions are expressed as functions of $x$ and $z_i$,  we replace the integration over $m^2_P$ with an integration over $x$
by using the relation in \cref{eq:relation_P2Q2x} and we turn the integration over the two-particle phase space
\begin{equation}
    \dd\phi_{2}(p_i, p_j;P) = (2\pi)^{D}\delta^{(D)}(P-p_{i}-p_{j})\dfrac{\dd ^{D}p_{i}}{(2\pi)^{D-1}}\delta_{+}(p^{2}_{i}-m^{2}_{i})\dfrac{\dd^{D} p_{j}}{(2\pi)^{D-1}}\delta_{+}(p^{2}_{j}-m^{2}_{j})
\end{equation}
into an integration over $z_i$. As a first step towards the parametrization through $z_i$, two Dirac delta functions are integrated out which gives
\begin{equation}
\begin{split}
    \dd\phi_{2}(p_i, p_j;P) = \frac{\dd^{D-1} p_{i}}{(2\pi)^{D-2}2E_{i}}\delta_{+}((P-p_i)^{2}-m^{2}_{j}). \label{eq:347x}
\end{split}    
\end{equation}
From now on, we will work in the c.m. frame of $p_i$ and $p_j$, i.e. in the rest frame of $P$, which sets the time and spatial components of $p_i$ and $p_j$ to the well-known expressions \cite{Byckling}
\begin{align}
&E_i = \frac{P^2 + m^2_i - m^2_j}{2\sqrt{P^2}}, \ \ \ \
 E_j = \frac{P^2 + m^2_j - m^2_i}{2\sqrt{P^2}}, \ \ \ \
 \abs{\vec{p}_i} = \abs{\vec{p}_j} = \frac{\sqrt{\lambda(P^2,m^2_i,m^2_j)}}{2\sqrt{P^2}} . \label{eq:pipjCM}
 \end{align}
 For the momentum $p_a$, we get
 \begin{align}
 &E_{a} = \frac{p_{a}\cdot P}{\sqrt{P^2}} = \frac{-\bar{Q}^2}{2x\sqrt{P^2} },\  \ \ \ \
\abs{\vec{p}_{a}}  = \sqrt{\frac{(p_{a}\cdot P)^2}{P^2}-m^2_{a}} = \frac{1}{2 \sqrt{P^2}} \frac{\sqrt{\lambda_{aj}}R(x)}{x}
\end{align}
where \cref{eq:relation_P2Q2x} was used to replace the product $p_a\cdot P$. The expressions in \cref{eq:pipjCM} can be used to write the remaining delta function in \cref{eq:347x} as a function of the norm of the momentum $\vec{p}_{i}$
\begin{equation}
    \delta_{+}(P^{2} -2P\cdot p_{i}+m^{2}_{i}-m^{2}_{j})= \frac{E_{i}}{2\sqrt{P^2}|\vec{p}_{i}|} \delta_{+}\left(\abs{\vec{p}_i} - \frac{1}{2\sqrt{P^2}}\sqrt{\lambda(P^2,m^2_i,m^2_j)}\right).
\end{equation} 
Inserting polar coordinates in $D-1$ dimensions
\begin{equation}
    \dd^{D-1}p_{i} = \dd\abs{\vec{p}_{i}} \  \abs{\vec{p}_{i}}^{D-2} \ \dd\Omega_{D-2}\  \dd\! \cos{\theta}\sin^{D-4}{\theta}
\end{equation}
allows to integrate out the remaining delta function and  the phase space measure becomes
\begin{equation}
    \dd\phi_{2}(p_i,p_j;P)
    = \dfrac{\dd\Omega_{D-2}}{(2\pi)^{D-2}}\dd\!\cos{\theta} \sin^{D-4}{\theta}\ \ \frac{1}{2}\left(  4 P^2\right)^{\frac{2-D}{2}} \lambda^{\frac{D-3}{2}}(P^2,m^2_i, m^2_j) \,
\end{equation}
where the angle $\theta$ is defined as the angle between $\vec{p}_a$ and $\vec{p}_i$, so that $\cos{\theta}$ is given by 
\begin{equation}
    \cos{\theta} = \frac{E_{i}E_{a}-p_{i}\cdot p_{a}}{|\vec p_{i}||\vec p_{a}|}.
\end{equation}
The integration over $\cos{\theta}$ can now be turned easily into an integration over the desired variable $z_i$ as $E_i$, $E_a$, $|\vec p_{i}|$ and $|\vec p_{a}|$ do not depend on $z_i$. In order to express $\sin{\theta}$ through $z_i$, the integration limits 
\begin{equation}
    z_\pm = \frac{E_i E_a \pm |\vec p_{i}||\vec p_{a}|}{P\cdot p_a},
\end{equation}
which are given in \cref{eq:int_limits_z} in terms of $x$ and $Q^2$ for $m_i=0$, can be used to write
\begin{align}
       &\sin^2{\theta} = (1-\cos{\theta})(1+\cos{\theta}) 
       = \left(\frac{P\cdot p_a}{|\vec p_{i}||\vec p_{a}|}\right)^2 (z_i - z_-)(z_+ - z_i).
\end{align}
As we assume rotational invariance of the squared matrix element around the axis given by $\vec{p}_a$, we can already perform the integration over the solid angle $\Omega_{D-2}$, so that we have
\begin{equation}
 \int\dd\Omega_{D-2} = \frac{2\pi^{\frac{D-2}{2}}}{\Gamma(\frac{D-2}{2})}.
\end{equation}
After defining the jacobian from the transition from $m^2_P$ to $x$ into the dipole phase space $\left[\dd p_i\left(Q^2,x,z_i\right)\right]$, one arrives at \cref{eq:FEIS_dip_PS}.

\subsection{Initial-state emitter and initial-state spectator}
\label{sec:PSIE-IS}
As in the previous section, the form of the measure $\left[\dd{p_i}(s,x,y)\right]$ is derived by considering a convolution of the form
\begin{multline}
\dd\phi_{m+1}( p_{i}, p_{k};p_a + p_b) 
 = \frac{\dd m^2_P}{2\pi} 
 \prod_{k}\frac{\dd[D]p_{k}}{(2\pi)^{D-1}}    \delta_{+}(p^{2}_{k}-m^{2}_{k})  (2\pi)^D\delta^{(D)}(p_{a}+p_{b}-p_{i}-\sum_{k}p_{k})  \dd\phi_2(p_i,P;p_a+p_b) \label{eq:IE-ISPS}
\end{multline}
where $m_{P}$ acts as the invariant mass of the momentum $P=p_a+p_b-p_i$.
By using the facts that the dipole momenta obey the mass-shell relations $\tilde{p}_k=m_k^2$ and momentum conservation $p_{a}+p_{b}-p_{i}-\sum_{k}p_{k}=\tilde{p}_{ai}+p_{b}-\sum_{k}\tilde{p}_{k}$ by construction and that a Lorentz transformation $\tilde{p}^\mu_k = {\Lambda^\mu}_\nu  p^\nu_k$ leaves the measure $\dd[D]p_{k}$ invariant, the remaining momentum integrations in \cref{eq:IE-ISPS} can be expressed through a $m$-particle phase space with initial momentum $\tilde{p}_{ai}+p_{b}$ and final momenta $\tilde{p}_k$:
\begin{equation}
    \dd\phi_{m+1}( p_{i}, p_{k};p_a + p_b) = \frac{\dd m^2_P}{2\pi} 
 \dd\phi_m(\tilde{p}_k;\tilde{p}_{ai}+p_b)  \dd\phi_2(p_i,P;p_a+p_b).
\end{equation}
Following the same line of thought as in \cref{sec:PSFE-IS} and working in the c.m. frame of $p_a$ and $p_b$, the integration over the two-particle phase space for $m_i=0$
\begin{equation}
    \dd\phi_2(P, p_i;p_a + p_b) = \dfrac{\dd^{D}p_i}{(2\pi)^{D-1}} \delta_{+}(p^2_i) \dfrac{\dd^{D}P}{(2\pi)^{D-1}} \delta_{+}(P^2-m^2_{P}) (2\pi)^{D}\delta^{(D)}(p_a + p_b - P - p_i) 
\end{equation}
can be turned into an integration over $y$
\begin{align}
   \dd\phi_2(P, p_i;p_a + p_b)  &= \frac{\dd\Omega_{D-2}}{2(2\pi)^{D-2}}\  \dd\abs{\vec{p_i}} \abs{\vec{p_i}}^{D-3}\  \dd\!\cos{\theta}\sin^{D-4}{\theta}\  \delta_{+}(s -2\abs{\vec{p_i}} \sqrt{s} - P^2) \nonumber \\
    &= \dfrac{\dd\Omega_{D-2}}{2(2\pi)^{D-2}} (2 \sqrt{s})^{2-D} (s-P^2)^{D-3}\  \dd\!\cos{\theta}\sin^{D-4}\theta \nonumber  \\
     &= \frac{\bar{s}^{1-2\varepsilon}}{2(4\pi)^{2-2\varepsilon}}\frac{(4s)^{-\varepsilon}}{\sqrt{\lambda_{ab}}^{1-2\varepsilon}}\  \dd\Omega_{D-2} \  \dd y \left[  (y_+ - y)(y - y_{-})   \right]^{-\varepsilon}.
\end{align}
The angle $\theta$ is defined as the angle between $\vec{p}_i$ and $\vec{p}_a$ and therefore determined through
\begin{equation}
    p_a \cdot p_i = \abs{\vec{p}_i} E_a - \abs{\vec{p}_a} \abs{\vec{p}_i} \cos{\theta}.
\end{equation}
A simple substitution from $m^2_P$ to $x$ via \cref{eq:stildeIEIS} yields the dipole phase space $ \left[\dd{p_i}(s,x,y)\right]$ given in \cref{eq:PSIEIS}. \\

\section{Integrals}
\label{sec:dipole_integrals}
The expansion in $\varepsilon$ of the integrals $I_1(z;\varepsilon)$ and $I_2(z;\varepsilon)$ up to $\order{\varepsilon}$ is obtained by inserting the ansatz 
\begin{equation}
u(z)=r(z) +\varepsilon  s(z)  +\order{\varepsilon^2}
\end{equation}
into the hypergeometric equation \cite{lebedev1972special}
\begin{equation}
z(1-z) u''(z)+(c-(a+b+1)z) u'(z)-a b  u(z)=0 \label{eq:ODE_2f1}
\end{equation}
whose general solution for the initial condition $u(0)=1$ is the hypergeometric function $u= \, _2F_1(a,b;c;z)$. Solving the resulting system of equations order by order while enforcing the boundary conditions $r(0)=1$ and $s(0)=0$ yields the functions $r(z)$ and $s(z)$.

For the computation of the integrals $\mathcal{I}_1\left(y_{0};\varepsilon\right)$ and $\mathcal{I}_2\left(y_{0};\varepsilon\right)$, the integral 
\begin{align}
\int_{0}^{\infty} \dd{t} t^{\alpha -1} \, _2F_1(a,b;c;-t)=\frac{\Gamma (\alpha ) \Gamma (c) \Gamma (a-\alpha ) \Gamma (b-\alpha )}{\Gamma (a) \Gamma (b) \Gamma (c-\alpha )} \label{eq:int_f21_gamma} 
\end{align}
is used. It can be computed by inserting the integral representation of the hypergeometric function followed by factorizing the double integral into two Euler-Beta functions
\begin{multline}
    \int_0^\infty \dd{t} \int_0^1 \dd{t'} t^{\alpha-1}  t'^{b-1}(1-t')^{c-b-1}(1+t t')^{-a} = \int_0^1 \dd{x} (1-x)^{\alpha-1} x^{a-\alpha-1} \int_0^1 \dd{t'} (1-t')^{c-b-1} t'^{b-\alpha-1} \\= \beta(\alpha,a-\alpha) \beta(c-b,b-\alpha)     
\end{multline}
through the substitution $x=\frac{1}{1+t t'}$. 
The remaining step for the computation of $\mathcal{I}_1\left(y_{0};\varepsilon\right)$ is to separate the integral into a part giving the divergences for $y \to 0$ and a finite part
\begin{equation}
\mathcal{I}_1(y_0;\varepsilon)=\beta(1-\varepsilon,1-\varepsilon)\left(\frac{1}{y^{1+\varepsilon}_0}\int_0^1 \dd{t} t^{\varepsilon} \, _2F_1\left(1,1-\varepsilon ;2-2 \varepsilon ;-\frac{t}{y_0}\right)
-\int_0^{\infty}\dd{t} t^{\varepsilon} \, _2F_1\left(1,1-\varepsilon ;2-2 \varepsilon ;-t\right)\right).
\end{equation}
The last part contains the divergent piece and is evaluated with the help of \cref{eq:int_f21_gamma} 
\begin{equation}
\int_0^{\infty}\dd{t} t^{\varepsilon} \, _2F_1(1,1-\varepsilon ;2-2 \varepsilon ;-t)=\frac{1}{2 \varepsilon ^2}-\frac{1}{\varepsilon }+\order{\varepsilon}
\end{equation}
whereas the first integral is finite and can be evaluated for $\varepsilon=0$ 
\begin{equation}
\int_0^1 \dd{t} t^{\varepsilon} \, _2F_1\left(1,1-\varepsilon ;2-2 \varepsilon ;-\frac{t}{y_0}\right)= - y_0 \operatorname{Li}_2\left(-\frac{1}{y_0}\right) + \order{\varepsilon}.
\end{equation}
The calculation of $\mathcal{I}_2\left(y_{0};\varepsilon\right)$ proceeds in an analogous way. 

\bibliography{references.bib}
\end{document}